\documentclass[%
reprint,
groupedaddress,
unsortedaddress,
nofootinbib,
 amsmath,amssymb,
 aps,
pra,
]{revtex4-2}

\usepackage[utf8]{inputenc}
\usepackage[english]{babel}
\usepackage{titlesec}
\usepackage{etoolbox}
\usepackage{dcolumn}
\usepackage{multirow}

\usepackage{graphicx} 
\usepackage{xcolor}
\usepackage{natbib}
\usepackage{braket}

\usepackage{siunitx}
\usepackage[all]{nowidow}
\usepackage[normalem]{ulem}
\usepackage{cancel}
\usepackage{bm}

\usepackage{float}
\usepackage{subcaption} 
\usepackage{enumitem}
\usepackage{hyperref}

\newcommand{\dd}{\mathrm{d}}

\newcommand{\inv}{^{-1}}

\DeclareSIUnit\bar{bar}
\DeclareSIUnit\angstrom{\text{\AA}}

\begin{document}

\makeatletter
\long\def\@makecaption#1#2{%
  \par\addvspace\abovecaptionskip
  \begingroup
  \leftskip=0pt
  \rightskip=0pt
  \parindent=0pt
  #1. #2\par
  \endgroup
}

\preprint{APS/123-QED}

\title{An Always-Accepting Algorithm for Transition Path Sampling}

\author{Magdalena H\"aupl 
}
\affiliation{Institute of Physics, University of Augsburg, Universit\"atsstraße 1, 86159 Augsburg, Germany}
\affiliation{Faculty of Physics, University of Vienna, 1090 Vienna, Austria}

\author{Sebastian Falkner 
}
\affiliation{Institute of Physics, University of Augsburg, Universit\"atsstraße 1, 86159 Augsburg, Germany}
\affiliation{Faculty of Physics, University of Vienna, 1090 Vienna, Austria}

\author{Peter G. Bolhuis 
}
\affiliation{Van ’t Hoff Institute for Molecular Sciences, University of Amsterdam, PO Box 94157, 1090GD Amsterdam, The Netherlands}

\author{Christoph Dellago 
}
\affiliation{Faculty of Physics, University of Vienna, 1090 Vienna, Austria}
\affiliation{Research Platform on Accelerating Photoreaction Discovery (ViRAPID), University of Vienna, 1090 Vienna, Austria}

\author{Alessandro Coretti 
}
\email{alessandro.coretti@univie.ac.at}
\affiliation{Faculty of Physics, University of Vienna, 1090 Vienna, Austria}

\date{\today}

\begin{abstract}
We present a one-way shooting algorithm for transition path sampling that accepts every proposed trajectory, yet samples the correct transition path ensemble for systems with overdamped stochastic dynamics. The method is based on two key elements: a procedure to propose trajectories that are always reactive, and a reweighting scheme that corrects for the bias introduced by always accepting the proposed paths. This approach significantly improves the efficiency of transition path sampling by eliminating the cost associated with generating trajectories that are then rejected. We demonstrate the performances of the algorithm by investigating the formation of CO$_2$ clathrate hydrates along different reaction mechanisms, showing that the increased efficiency allows proper sampling of the formation of crystalline hydrates at temperatures and pressures that are difficult to access with conventional schemes.
\end{abstract}

\maketitle

\section{Introduction}
Rare events are ubiquitous in many areas of science, from nucleation processes to biomolecular reorganizations~\cite{pan_dynamics_2004,juraszek_sampling_2006,okazaki_mechanism_2019,angiolari_electrically_2025,falkner_kinetic_2021}. The use of computational methods and numerical simulations to investigate these processes is pivotal as it gives insight into the microscopic mechanisms at the basis of the event. Standard equilibrium simulation techniques struggle since, although the event is rare by definition, it typically occurs rapidly, requiring high time and space resolution to resolve all important details at a microscopic level. To overcome these issues, computational methods have been introduced to steer the sampling towards the important barrier regions, for example by introducing explicit biasing potentials~\cite{torrie_nonphysical_1977,laio_escaping_2002}. Among these, transition path sampling (TPS)~\cite{dellago_transition_1998, bolhuis_trowing_2002} has emerged as a powerful technique, mostly because it does not require knowledge of a reaction coordinate to describe the transition and preserves the natural dynamics of the process. Indeed, the bias towards the transition region is not obtained through an additional term in the potential energy, but via a specific Markov Chain Monte Carlo algorithm that explores the space of reactive paths. The algorithm iteratively proposes a new path starting from an old one and accepts or rejects the generated path according to an acceptance criterion derived from the condition of detailed balance. In this way, the method samples the true transition path ensemble that is compatible with the equilibrium distribution of the system. 

Different versions of TPS have been proposed in the literature~\cite{bolhuis_trowing_2002,peters_obtaining_2006,peters_recent_2010,menzl_s-shooting_2016,jung_transition_2017,bolhuis_transition_2021,falkner_conditioning_2023,falkner_revisiting_2025,falkner_enhanced_2024,grunwald_precision_2008,lazzeri_molecular_2023}, but the core of most popular implementations relies on the so-called shooting move. Here, a configuration is randomly selected on the old path, possibly perturbed, and a trajectory is initiated from this shooting point. For the generated trajectory to be accepted, a necessary (but in general not sufficient) condition is that the generated path is reactive, i.e., that it connects the stable states of the reaction. While this prescription forms the foundation for accurately capturing the true transition dynamics of the system, it also represents a significant computational bottleneck of the method: each proposal move consists of the generation of an entire new trajectory, a process that can be computationally expensive for systems with many degrees of freedom and long transition times. This numerical cost must already be paid before the new path is tested against the acceptance criterion, as there is in general no way to know whether the trajectory will be reactive before generating it. Consequently, rejection of proposed paths may lead to substantial wasted computational resources, significantly limiting the overall efficiency of TPS.

Early attempts to mitigate the problem of low acceptance include the use of the (flexible length) one-way shooting algorithm~\cite{dellago_efficient_1998,Dellago2002, bolhuis_transition_2003,bolhuis_transition-path_2003,juraszek_sampling_2006}, in which only one segment (forward or backward) of the trajectory was renewed in a trial move.  While  boosting the acceptance ratio considerably, it came at the cost of  requiring  multiple attempts to achieve full decorrelation of a path.

Here, we present a flexible length TPS algorithm, applicable if inertial effects can be neglected on the time scale of the transition, that 
always produces reactive paths. However, this alone does not guarantee that every generated trajectory will be accepted and included in the transition path ensemble. Acceptance or rejection remains governed by the Metropolis criterion, which ensures that the relative statistical weights of paths are preserved in the transition path ensemble. In order to accept every trajectory produced, we then propose a reweighting technique that assigns the correct weights to each proposed trajectory a posteriori, ensuring that the correct ensemble is sampled and any bias is removed, avoiding at the same time any computational waste arising from rejecting trajectories that have already been generated.
While the benefits are system dependent and influenced by the underlying path distribution, the proposed technique ultimately enhances the overall performance of the method by improving exploration of transition regions and facilitating switching between reaction channels.

\section{Theory}
To define the problem mathematically, we identify two disjoint regions in  configuration space, which we call $\mathrm{A}$ and $\mathrm{B}$. We define a path $X$ of length $L$ that connects $\mathrm{A}$ and $\mathrm{B}$ as a collection of ordered configurations $X(L) = \{x_0,\dots,x_{L-1}\}$ separated by the timestep $\delta t$, where
$x_i$
is a point in the configuration space of the system such that $x_0\in\mathrm{A}$ and $x_{L-1}\in\mathrm{B}$. The probability distribution of reactive paths of length $L$ can be written as
\begin{equation}
\label{eq:path_AB}
P_{\mathrm{AB}}[X(L)] = Z_{\mathrm{AB}}\inv H_\mathrm{AB}[X(L)]P[X(L)],
\end{equation}
where the probability $P[X(L)]$ of any path $X(L)$, i.e., not limited to the reactive ones, can be written as
\begin{equation}
\label{eq:path}
P[X(L)] = \rho_{\text{eq}}(x_0)\prod_{i=0}^{L-2}p(x_i\to x_{i+1}).
\end{equation}
The transition kernel $p(x_i\to x_{i+1})$ is the conditional probability that the system is in $x_{i+1}$ at time $t = (i+1)\delta t$ given that it was in $x_i$ at time $t = i\delta t$. In what follows, we assume this kernel to be non-deterministic, i.e. stochastic. This can be (and, in fact, is) routinely realized by coupling the system with a stochastic thermostat (e.g., a Langevin thermostat)~\cite{dellago_efficient_1998}. In Eq.~\eqref{eq:path_AB}, $\rho_{\text{eq}}(x_0)$ is the equilibrium distribution of the initial point, and the function $H_\mathrm{AB}[X(L)]$ is unity if the path is reactive, namely it connects $\mathrm{A}$ and $\mathrm{B}$, with its initial and final points lying in different states and with no other points in any of the stable states; otherwise it vanishes. The normalization factor $Z_{\mathrm{AB}}$ is the partition function of the ensemble of reactive trajectories of any length bigger or equal than 2 frames and is defined by
\begin{equation}
Z_{\mathrm{AB}} = \sum_L \int \dd x_0\dots \dd x_{L-1} H_\mathrm{AB}[X(L)]P[X(L)].
\end{equation}
where the sum extends to infinity in principle but converges fast enough to be truncated at some $L_{\text{max}}$.
Equation~\eqref{eq:path_AB} is the target probability of the proposed sampling algorithm. 

\subsection{Always-Reactive Algorithm Transition Path Sampling  (ARA-TPS)}
\label{sec:always-reactive}
To achieve efficient sampling of the transition path ensemble $P_{\mathrm{AB}}[X(L)]$, we propose the following steps to generate a new path from an old one:
\begin{enumerate}[label=\textbf{Step \arabic*:}, ref=\textbf{Step~\arabic*}, leftmargin=*, itemindent=2.5em]
\item \label{itm:sp_sel} Starting from an old path $X(L)$ that starts in $\mathrm{A}$ and ends in $\mathrm{B}$, select a shooting point $x_s\in X(L)$ according to some selection probability $p_{\text{sel}}(x_s|X)$.
\item \label{itm:split} Divide the old path $X(L)$ in $X_{\text{bw}}= \{x_0,\dots,x_{s-1}\}$ and $X_{\text{fw}} = \{x_{s+1},\dots,x_{L-1}\}$ (note that the shooting point $x_s$ is excluded in both cases).
\item \label{itm:shoot} Start a trajectory from $x_s$ (without perturbing it) according to the rules of the underlying dynamics until it reaches one of the stable states, either $\mathrm{A}$ or $\mathrm{B}$ and call it $X'_{\text{new}} = \{x'_0,\dots,x'_{\ell'-1}\}$ where $\ell'$ denotes the length of the newly generated segment. Then
\begin{enumerate}[label=\textbf{(\alph*):}, ref=\textbf{\theenumi\alph*}, leftmargin=*]
\item \label{itm:fwd_move} If $X'_{\text{new}}$ ends in $\mathrm{B}$, reindex $X'_{\text{new}}$ to $X'_{\text{fw}} = \{x'_{s'},\dots,x'_{L'-1}\}$, and set the new path $X'(L')$ to be the ordered union of $X_{\text{bw}}$ and $X'_{\text{fw}}$. This yields $X'(L') = \{x_0,\dots,x_{s-1}, x'_{s'},\dots,x'_{L'-1}\}$ with $x'_{s'} = x_s$, $s'=s$ and $L' = s + \ell'$.
\item \label{itm:bwd_move} If $X'_{\text{new}}$ ends in $\mathrm{A}$, reverse the order of points in $X'_{\text{new}}$ and relabel them to obtain $X'_{\text{bw}} = \{x'_{0}, x'_{1}, \dots, x'_{s'}\}$. Then, reindex the points on $X_{\text{fw}}$ as $X_{\text{fw}} = \{x_{s'+1},\dots,x_{L'-1}\}$. The new path $X'(L')$ is then the ordered union of $X'_{\text{bw}}$ and $ X_{\text{fw}}$ which yields $X'(L') = \{x'_0,\dots,x'_{s'}, x_{s'+1},\dots,x_{L'-1}\}$ where $L' = s'+1 + (L'-s'-1) = \ell' + (L-s-1)$ since $s'+1=\ell'$ and $L'-s'=L-s$. Note that in this case, $s'\neq s$ but $x'_{s'} = x_s$.
\end{enumerate}
\item \label{itm:acc_rej} Accept or reject the new path according to a criterion that satisfies detailed balance.
\end{enumerate}

\begin{figure}[htbp]
\includegraphics[width=\columnwidth]{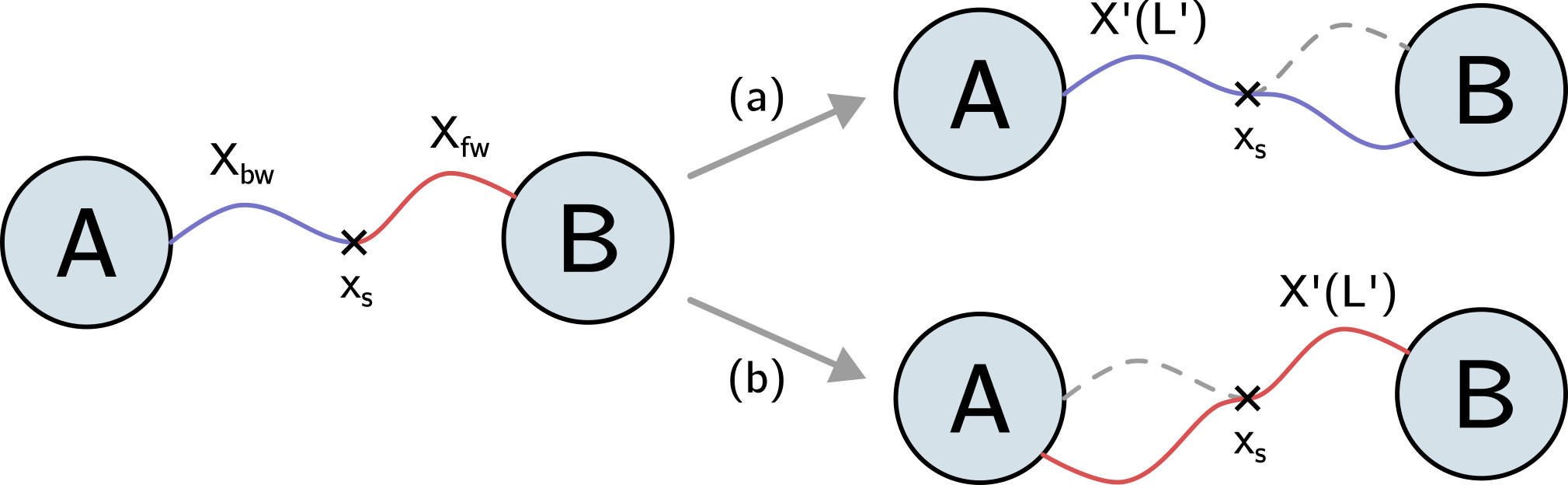}
\caption{\label{fig:ara-tps} Schematic representation of the always-reactive path generation algorithm. The old path is divided at $x_s$ and a new trajectory $X'_{\text{new}}$ is initiated from there. If $X'_{\text{new}}$ ends up in $\mathrm{B}$ (\ref{itm:fwd_move}) a new reactive trajectory $X'(L')$ is generated by reindexing $X'_{\text{new}}$ to $X'_{\text{fw}}$ and by adding it to the backward segment ($X_\text{bw}$ in blue). If $X'_{\text{new}}$ ends up in $\mathrm{A}$ (\ref{itm:bwd_move}), a new reactive trajectory $X'(L')$ is generated by reversing and relabeling $X'_{\text{new}}$, now denoted $X'_{\text{bw}}$, and appending it to the forward segment of the reindexed existing path ($X_\text{fw}$ in red). Dashed lines represent the part of the old path $X$ that is changed during the generation process.}
\end{figure}

This algorithm (schematically represented in Fig.~\ref{fig:ara-tps}) always produces reactive paths that start in $\mathrm{A}$ and end in $\mathrm{B}$, i.e., by construction, one always has $H_\mathrm{AB}[X']=1$.\footnote{Note that there exists a completely equivalent algorithm where in point~\ref{itm:bwd_move}, instead of reversing $X'_{\text{new}}$, $X_{\text{fw}}$ is reversed and denoted $X_{\text{bw}}$. The new path is then given by the ordered union of $X_{\text{bw}}$ and $X'_{\text{fw}} \equiv X'_{\text{new}}$. At difference with the algorithm outlined above --- which preserves the initial and final states of the original path along the simulation --- this variant produces paths that connect the stable states in any order ($\mathrm{A}$ to $\mathrm{B}$ and viceversa). We choose to focus on the first variant as it more closely matches the property of standard one-way shooting (which, in its original form~\cite{bolhuis_transition_2003,bolhuis_transition-path_2003}, also only produces paths that preserve initial and final states).} For this algorithm to produce physically meaningful trajectories, the assumption of overdamped stochastic dynamics is crucial. In this way, continuity in the momenta, which would be violated by the algorithm in~\ref{itm:bwd_move}, can be relaxed. Continuity in the coordinates, on the other hand, is guaranteed by using an unperturbed configuration on an old path as shooting point, a condition that is allowed by the choice of a stochastic transition kernel.

The acceptance criterion in~\ref{itm:acc_rej} of the scheme above can be determined from the detailed balance condition, 
which, for sampling the transition path ensemble, reads
\begin{equation}
\label{eq:det_bal_start}
\begin{split}
P_{\mathrm{AB}}[X(L)]&P[X(L)\to X'(L')] \\
&= P_{\mathrm{AB}}[X'(L')]P[X'(L')\to X(L)],
\end{split}
\end{equation}
where $P[X(L)\to X'(L')]$ is the transition probability from $X(L)$ to $X'(L')$. This probability can be factorized as $P[X(L)\to X'(L')] = P_{\text{gen}}[X(L)\to X'(L')]P_{\text{acc}}[X(L)\to X'(L')]$ and, removing the dependence on the path length for simplicity, we obtain
\begin{equation}
\label{eq:det_bal}
\begin{split}
P_{\mathrm{AB}}&[X]P_{\text{gen}}[X\to X']P_{\text{acc}}[X\to X'] \\
&= P_{\mathrm{AB}}[X']P_{\text{gen}}[X'\to X]P_{\text{acc}}[X'\to X]. \\
\end{split}
\end{equation}
\vspace{0.\baselineskip}

\noindent Equation~\eqref{eq:det_bal} provides the ratio of acceptance probabilities
\begin{equation}
\label{eq:acc_det_bal}
\frac{P_{\text{acc}}[X\to X']}{P_{\text{acc}}[X'\to X]} = \frac{P_{\mathrm{AB}}[X']P_{\text{gen}}[X'\to X]}{P_{\mathrm{AB}}[X]P_{\text{gen}}[X\to X']},
\end{equation}
which can be satisfied using the Metropolis-Hastings acceptance criterion
\begin{equation}
\label{eq:MH_acc_I}
P_{\text{acc}}[X\to X'] = \min\left\{1, \frac{P_{\mathrm{AB}}[X']P_{\text{gen}}[X'\to X]}{P_{\mathrm{AB}}[X]P_{\text{gen}}[X\to X']}\right\}.
\end{equation}
Equation~\eqref{eq:MH_acc_I} can be further simplified by using Eq.~\eqref{eq:path_AB} and by noting that the old and new paths $X$ and $X'$ are reactive by construction, i.e. $H_\mathrm{AB}[X] = 1$ and $H_\mathrm{AB}[X'] = 1$. One finally obtains
\begin{equation}
\label{eq:MH_acc_II}
P_{\text{acc}}[X\to X'] = \min\left\{1, \frac{P[X']P_{\text{gen}}[X'\to X]}{P[X]P_{\text{gen}}[X\to X']}\right\},
\end{equation}
where the partition functions $Z_{\mathrm{AB}}$ simplify in the fraction.

Equation~\eqref{eq:MH_acc_II} provides a practical way to accept or reject paths generated by the algorithm discussed above once the generation probability has been specified. According to~\ref{itm:sp_sel} and~\ref{itm:shoot} in the algorithm above, we have
\begin{equation}
\label{eq:p_gen}
P_{\text{gen}}[X\to X'] = p_{\text{sel}}(x_s|X)\prod_{j=0}^{\ell'-2}p(x'_j\to x'_{j+1}),
\end{equation}
with $x_0'=x_s$. In~\ref{itm:fwd_move}, substituting Eqs.~\eqref{eq:path} and~\eqref{eq:p_gen} in Eq.~\eqref{eq:MH_acc_II}, trivial cancellations yield
\begin{equation}
\label{eq:acc_final}
P_{\text{acc}}[X\to X'] = \min\left\{1, \frac{p_{\text{sel}}(x'_{s'}|X')}{p_{\text{sel}}(x_s|X)}\right\}.
\end{equation}
To prove that this is also true for~\ref{itm:bwd_move}, we expand the fraction in Eq.~\eqref{eq:MH_acc_II} using Eqs.~\eqref{eq:path} and~\eqref{eq:p_gen}. We obtain
{
\setlength{\abovedisplayskip}{15pt}
\setlength{\belowdisplayskip}{15pt}
\setlength{\abovedisplayshortskip}{15pt}
\setlength{\belowdisplayshortskip}{15pt}

\begin{widetext}
\begin{equation}
\label{eq:acc_B2A}
\begin{split}
\frac{P[X']P_{\text{gen}}[X'\to X]}{P[X]P_{\text{gen}}[X\to X']} = 
\frac{\rho_{\text{eq}}(x'_0)\prod_{i=0}^{s'-1}p(x'_i\to x'_{i+1})\prod_{i=s'}^{L'-2}p(x'_i\to x'_{i+1})}{\rho_{\text{eq}}(x_0)\prod_{i=0}^{s-1}p(x_i\to x_{i+1})\prod_{i=s}^{L-2}p(x_i\to x_{i+1})}\frac{p_{\text{sel}}(x'_{s'}|X')\rho_{\text{eq}}(x_s)\prod_{j=0}^{\ell-2}p(x_j\to x_{j+1})}{p_{\text{sel}}(x_s|X)\rho_{\text{eq}}(x'_{s'})\prod_{j=0}^{\ell'-2}p(x'_j\to x'_{j+1})},
\end{split}
\end{equation}
\end{widetext}
}
\noindent where we have split the path probability of Eq.~\eqref{eq:path} in correspondence of the shooting points $x'_{s'} = x_s$ in the numerator and in the denominator, and we have multiplied and divided by $\rho_{\text{eq}}(x_s) = \rho_{\text{eq}}(x'_{s'})$. This splitting corresponds to point~\ref{itm:split} of the algorithm. The generated path can now be reversed via the microscopic reversibility condition, which takes the form 
$\rho_{\text{eq}}(x)p(x\to y) = \rho_{\text{eq}}(y)p(y \to x)$. This equation can be inserted in Eq.~\eqref{eq:acc_B2A} to reverse the generated paths, so to get
\begin{equation}
\label{eq:app_MR}
\begin{aligned}
\rho_{\text{eq}}(x_s)\prod_{j=0}^{\ell-2}p(x_j\to x_{j+1}) &= \rho_{\text{eq}}(x_{\ell-1})\prod_{j=1}^{\ell-1}p(x_{j}\to x_{j-1}), \\
\rho_{\text{eq}}(x'_{s'})\prod_{j=0}^{\ell'-2}p(x'_j\to x'_{j+1}) &= \rho_{\text{eq}}(x'_{\ell'-1})\prod_{j=1}^{\ell'-1}p(x'_{j}\to x'_{j-1}).
\end{aligned}
\end{equation}
Reordering the terms in the product on the RHS of Eq.~\eqref{eq:app_MR} 
we obtain
\begin{equation}
\begin{aligned}
\rho_{\text{eq}}(x_s)\prod_{j=0}^{\ell-2}p(x_j\to x_{j+1}) &= \rho_{\text{eq}}(x_0)\prod_{j=0}^{s-1}p(x_{j}\to x_{j+1}),\\
\rho_{\text{eq}}(x'_{s'})\prod_{j=0}^{\ell'-2}p(x'_j\to x'_{j+1}) &= \rho_{\text{eq}}(x'_0)\prod_{j=0}^{s'-1}p(x'_{j}\to x'_{j+1}),\\
\end{aligned}
\end{equation}
where we recall that, in the present notation, $x_{\ell-1}\in\mathrm{A}$ and $x_{j=0} = x_s$ and the same for $X'$. This corresponds to~\ref{itm:bwd_move} of the algorithm. Substituting these expressions into Eq.~\eqref{eq:acc_B2A} and recalling that the last part of the path (from the configuration index $s^{(}{}'{}^{)}$ to $L^{(}{}'{}^{)} - 1$ in $X^{(}{}'{}^{)}$) is shared between $X$ and $X'$, we obtain
\[
\frac{P[X']P_{\text{gen}}[X'\to X]}{P[X]P_{\text{gen}}[X\to X']} = 
\frac{p_{\text{sel}}(x'_{s'}|X')}{p_{\text{sel}}(x_s|X)},
\]
and thus Eq.~\eqref{eq:acc_final} is also proved for~\ref{itm:bwd_move}. Equation~\eqref{eq:acc_final} is the acceptance criterion normally used in TPS and has already been derived elsewhere for different algorithms~\cite{jung_transition_2017}. This acceptance criterion shows that even though the algorithm proposed always produces reactive paths, their acceptance is still bounded by the ratio between the shooting point selection probabilities. This ratio takes different forms depending on the choice of $p_{\text{sel}}(x_s|X)$: for example,
it reduces to the ratio between the path lengths $L$ and $L'$ in the case of uniform selection probability or to the ratio of the weight normalizations in the case of shooting range (SR)~\cite{jung_transition_2017} or a Gaussian selector~\cite{Bolhuis_2010}.

\subsection{Always-Accepting Algorithm for Transition Path Sampling (AAA-TPS)}

Having provided a method that always produces reactive paths, it is tempting to completely eliminate waste of computational resources by always accepting the proposed trajectories produced by ARA-TPS. Doing this, results in the algorithm sampling the biased distribution~\cite{falkner_revisiting_2025}
$\widetilde{P}_{\mathrm{AB}}[X(L)] = \Omega[X(L)]P_{\mathrm{AB}}[X(L)]$, where $\Omega[X(L)]$ are statistical weights attached to each path sampled by the algorithm. Dropping once again the dependence on the path length for brevity, the detailed balance condition for this auxiliary transition path ensemble reads
\begin{equation}
\begin{split}
\widetilde{P}_{\mathrm{AB}}[X]&P[X\to X'] = \widetilde{P}_{\mathrm{AB}}[X']P[X'\to X],
\end{split}
\end{equation}
which, through the same steps that take Eq.~\eqref{eq:det_bal_start} to Eq.~\eqref{eq:acc_det_bal}, yields the ratio of acceptance probabilities 
\begin{equation}
\frac{P_{\text{acc}}[X\to X']}{P_{\text{acc}}[X'\to X]} = \frac{\widetilde{P}_{\mathrm{AB}}[X']P_{\text{gen}}[X'\to X]}{\widetilde{P}_{\mathrm{AB}}[X]P_{\text{gen}}[X\to X']}.
\end{equation}
By imposing the acceptance of every path generated by the algorithm, namely $P_{\text{acc}}[X\to X'] = 1 \ \forall \ X, X'$, we can derive an expression for the weights $\Omega[X]$. To this end, we substitute $\widetilde{P}_{\mathrm{AB}}[X] = \Omega[X]P_{\mathrm{AB}}[X]$ and simplify the normalization factors as well as the characteristic functions, which are equal to unity for the generation algorithm described in Sec.~\ref{sec:always-reactive}. In this way, we obtain
\begin{equation}
1 = \frac{\Omega[X']P[X']P_{\text{gen}}[X'\to X]}{\Omega[X]P[X]P_{\text{gen}}[X\to X']},
\end{equation}
which yields the following expression for the ratio of the weights, 
\begin{equation}
\label{eq:weights_ratio_start}
\frac{\Omega[X']}{\Omega[X]} = \frac{P[X]P_{\text{gen}}[X\to X']}{P[X']P_{\text{gen}}[X'\to X]}.
\end{equation}
For a set of statistical weights $\Omega$ that satisfy Eq.~\eqref{eq:weights_ratio_start}, detailed balance is obeyed. Hence $\widetilde{P}_{\mathrm{AB}}[X] = \Omega[X(L)] \times P_{\mathrm{AB}}[X(L)]$, if it exists, is the stationary distribution sampled by the Markov Chain. The RHS of Eq.~\eqref{eq:weights_ratio_start} is the inverse of the ratio appearing in the second term of the minimum operator for the Metropolis-Hastings criterion in Eq.~\eqref{eq:MH_acc_II}. Assuming the same path generation rule for the proposed algorithm, the same steps that lead from Eq.~\eqref{eq:MH_acc_II} to Eq.~\eqref{eq:acc_final} yield
\begin{equation}
\label{eq:weights_ratio_middle}
\frac{\Omega[X']}{\Omega[X]} = \frac{p_{\text{sel}}(x_s|X)}{p_{\text{sel}}(x'_{s'}|X')}.
\end{equation}
Note that, to preserve continuity of coordinates and generate physically meaningful trajectories with the proposed algorithm, one must always have $x_s = x'_{s'}$. For this reason, any selection probability of the form~\cite{jung_transition_2017}
\begin{equation}
\label{eq:psel_w}
p_{\text{sel}}(x_s|X) = \frac{\omega(x_s)}{\sum_{x_i\in X}\omega(x_i)},
\end{equation}
with $\omega(x) \ge 0$ for all $x$ (and $> 0$ for the shooting points), yields the expression
\begin{equation}
\label{eq:weights_ratio_final}
\frac{\Omega[X']}{\Omega[X]} = \frac{\sum_{x'_i\in X'}\omega(x'_i)}{\sum_{x_i\in X}\omega(x_i)},
\end{equation}
which allows us to associate the (unnormalized) weight $\Omega[X] = \sum_{x_i\in X}\omega(x_i)$ to the path $X$ when sampling with unit acceptance. Note that the $\omega(x)$ appearing in Eq.~\eqref{eq:weights_ratio_final} are the unnormalized weights associated to configurations on a path obtained from the selection probabilities (or, in other words, the weight $\Omega[X]$ associated to the path is the normalization constant in Eq.~\eqref{eq:psel_w}). As expected, $\Omega[X]$ depends on the path and not on the shooting point, as each path can be generated from multiple configurations. To recover the correct transition path ensemble from $\widetilde{P}_{\mathrm{AB}}[X(L)]$ each generated path must be resampled or reweighted with weight $\Omega\inv[X]$. For a uniform shooting point selection probability, the weight associated with each path is given by $\Omega[X] = L(X)$, as already noted in Ref.~\citenum{falkner_revisiting_2025}.

We emphasize that, at this stage, the two methods introduced, the Always-Reactive Algorithm (ARA-TPS) and the Always-Accepting Algorithm (AAA-TPS) may be employed independently of one another. 
In particular, it is both possible and potentially beneficial to employ an a posteriori reweighting technique in combination with standard one-way shooting (or even two-way shooting), for instance when some of the dynamical conditions (e.g., negligible inertial effects) required by the always-reactive algorithm are not satisfied.
When the conditions for applying the path-generation algorithm described in Sec.~\ref{sec:always-reactive} are met, however, there appears to be little justification for using standard one-way shooting in place of ARA-TPS, as the latter immediately yields a twofold increase in the acceptance rate (see also Secs.~\ref{sec:num_res} and~\ref{sec:disc} below). By contrast, criteria for determining when a posteriori reweighting is more efficient than explicitly testing the Metropolis criterion and rejecting a subset of the generated paths have not yet been addressed. Reweighting procedures between probability distributions are well established in atomistic simulations~\cite{morawietz_how_2016, shirts_reweighting_2017, henin_enhanced_2022, montero_de_hijes_density_2024} and path sampling methods~\cite{rogal_reweighted_2010, zhang_weighted_2010, keller_dynamical_2024, singh_variational_2025, erp_generalized_2026}, and their efficiency is known to depend on the degree of overlap between the original and target probability densities~\cite{lu_improving_2004,bennett_efficient_1976,shirts_statistically_2008,kumar_weighted_1992,zwanzig_high-temperature_1954}. In recent years, the Kish effective sample size (ESS)~\cite{Kish1966-hz} has emerged as a common metric to quantify this efficiency~\cite{coretti_learning_2025,jung_normalizing_2024,wirnsberger_estimating_2023,wang_energy_2026,schebek_efficient_2024}, which, in the current scenario, is defined as
\begin{equation}
\label{eq:ess}
\text{ESS} = \frac{\left(\sum_{i=1}^N\Omega\inv[X_i]\right)^2}{\sum_{i=1}^N\left(\Omega\inv[X_i]\right)^2},
\end{equation}
where $N$ is the number of paths involved in the reweighting procedure and $\Omega\inv[X_i]$ are the corresponding path weights obtained from Eq.~\eqref{eq:weights_ratio_final} which yield the correct transition path ensemble distribution $P_{\mathrm{AB}}[X(L)]$. The ESS can have values between $1$ (one weight dominates) and $N$ (all weights are equal). 
An analogous concept of effective sample size can be introduced in the context of acceptance–rejection methods by considering the acceptance rate $r_{\text{acc}}$, which is typically estimated as the fraction of accepted samples over the total number of proposals. A rough estimate of the effective sample size for a Monte Carlo method with a Metropolis acceptance criterion is therefore given by $N r_{\text{acc}}$ where $N$ is again the total number of proposed samples. A first criterion to determine whether it is more efficient to use ARA-TPS compared to AAA-TPS is then represented by
\begin{equation}
Nr_{\text{acc}} > \mathrm{ESS}.
\end{equation}
This equation, however, neglects correlation effects, whose magnitudes depend on the observable considered. These correlations can be quantified, for a given observable, through the integral of its normalized autocorrelation function~\cite{zwanzig_statistical_1969}. A more robust estimate of the efficiency associated with the two methods can then be obtained by analyzing a posteriori the autocorrelation functions of the particular observables investigated. An example in this sense is provided in Sec.~\ref{subsec:2d_models}.

\section{Numerical Results}
\label{sec:num_res}
\subsection{Two-dimensional models}
\label{subsec:2d_models}
\begin{figure*}[htbp]
\includegraphics[width=\textwidth]{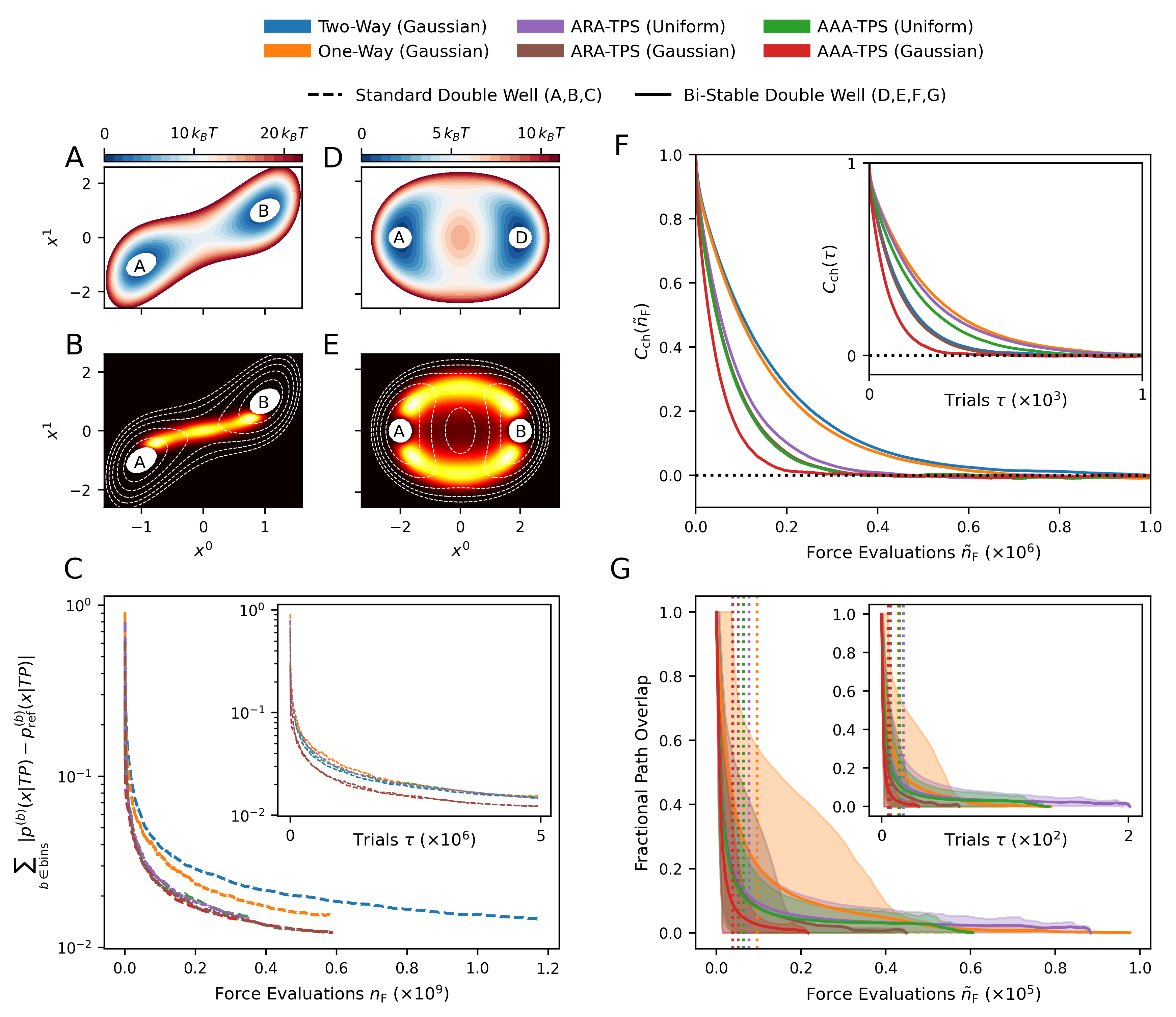}
\caption{\label{fig:2d_models} Results of the numerical simulations for the two-dimensional models. \textbf{Standard double well:} (A) Potential Energy Surface (PES) and stable state definitions. (B) $p(x|\text{TP})$ obtained via the normalized histogram of the configurations on transition paths. (C) Sum of bin-by-bin absolute difference between $p(x|\text{TP})$ obtained with the methods investigated and the reference transition path ensemble as a function of the number of force evaluations (main) and the number of TPS trials (inset). \textbf{Bi-stable double well:} (D) and (E) are the same as (A) and (B) for this PES. (F) Autocorrelation function of transition channel: to each transition path in the ensemble $X(L)$ a value of $1$ or $-1$ is assigned depending on whether the average $x^1$ on the trajectory is positive or negative; the autocorrelation function of this quantity is plotted as a function of the number of force evaluations (main) and of the number of TPS trials (inset). (G) Average fractional overlap between successive paths as a function of the number of force evaluations (main) and the number TPS trials (inset). For each trajectory in the ensemble, the fractional overlap with successive trajectories is computed until no configurations are shared. The vertical dotted lines represent how many force evaluations (main) or TPS trials (inset) are needed, on average, to obtain a completely new trajectory. Shaded areas represent the 10th and 90th percentiles of the distribution of path overlaps.}
\end{figure*}
We first assess the correctness of the proposed algorithm by investigating the convergence of $p(x|\text{TP})$, namely the probability density of a configuration $x$ given that it belongs to a transition path, for a simple double well potential with a $10 \, k_BT$ barrier as depicted in Fig.~\ref{fig:2d_models}A (details on the potential energy and on the dynamics are given in the Supplementary Material). 
In particular, we compute the sum of the absolute bin-by-bin difference between a reference $p_{\text{ref}}(x|\text{TP})$ and the one calculated with the algorithms introduced as a function of the number of transition paths (Fig.~\ref{fig:2d_models}C). For the reference transition path ensemble (Fig.~\ref{fig:2d_models}B), we use the normalized histogram of the configurations on transition paths from a long two-way shooting TPS simulation with uniform selection probability. We benchmark ARA-TPS and AAA-TPS with two different shooting point selection probabilities: With the definition given in Eq.~\eqref{eq:psel_w}, in one case we use uniform weights $\omega(x) = 1$; in a second test, we use Gaussian weights $\omega(x) = \exp[-k(c(x) - c_{\text{ref}})^2]$ with $c(x) \equiv x^0+x^1$ as a collective variable, $c_{\text{ref}} = 0$ as bias center and $k=12.5$ in reduced units as a force constant. For comparison, we also show the results obtained with standard two-way~\cite{Dellago2002} and one-way shooting~\cite{bolhuis_transition-path_2003} using the same Gaussian selection probability. All the details of the simulations are reported in the Supplementary Material. The algorithms introduced, with both uniform and Gaussian selection probabilities, show better performances in converging to the reference ensemble when computed in terms of the number of force evaluations. In terms of the number of TPS trials, the Gaussian selection probability produces better convergence with respect to all other methods. 
\begin{table}[btp]
\caption{\label{tab:stdw_comp_ARAvsAAA}
Acceptance rate $r_{\text{acc}}$ and Effective Sample Size (ESS) normalized by the number of samples $N$ for the methods investigated. Simulations performed on a standard double well PES (Fig.~\ref{fig:2d_models}A).}
\begin{ruledtabular}
\begin{tabular}{ccc}
\multirow{2}{*}{TPS Method} & \multirow{2}{*}{$r_{\text{acc}}$} & \multirow{2}{*}{$\text{ESS}/N$} \\ \\
\hline
Two-Way Shooting\footnotemark[1]& $0.25$& $-$\\
One-Way Shooting\footnotemark[1]& $0.40$& $-$\\
Always-Reactive Algorithm\footnotemark[1]& $0.80$& $-$\\
Always-Reactive Algorithm\footnotemark[2]& $0.92$& $-$  \\
Always-Accepting Algorithm\footnotemark[1]& $-$& $0.72$\\
Always-Accepting Algorithm\footnotemark[2]& $-$& $0.91$\\
\end{tabular}
\end{ruledtabular}
\footnotetext[1]{Gaussian shooting point selection probability.}
\footnotetext[2]{Uniform shooting point selection probability.}
\end{table}
We observe, for this specific potential, an almost perfect superposition of the results for ARA-TPS and AAA-TPS with the respective selection probabilities. The acceptance rate and effective sample size are reported in Table~\ref{tab:stdw_comp_ARAvsAAA}. This example shows how the condition $Nr_{\text{acc}} > \mathrm{ESS}$ is not sufficient to guarantee a better efficiency of ARA-TPS with respect to AAA-TPS, at least for the observable considered in Fig.~\ref{fig:2d_models}C. In Tab.~\ref{tab:stdw_comp_ARAvsAAA}, we find that the acceptance rate of standard one-way shooting is exactly half of ARA-TPS for the same shooting point selection probability. This behavior is hardly surprising, since in the overdamped regime the a priori choice of a ``direction'' has no effect on the dynamics, and the only effect is the rejection of the half of the trajectories that end up in the ``wrong'' state.

A more challenging problem in two dimensions is represented by the bi-stable double well shown in Fig.~\ref{fig:2d_models}D (details of the potential energy surface in the Supplementary Material).
In this case, there are two possible pathways, symmetric in the $x^{1}$ direction, that connect states $\mathrm{A}$ and $\mathrm{B}$ (Fig.~\ref{fig:2d_models}E) and that are separated by an energy barrier of approximately $10\,k_BT$. The sampling can therefore be trapped in one of the channels, and correlations between subsequently sampled paths are expected to play a bigger role.
For this system, we investigate the efficiency of the proposed methods, again with Gaussian (same parameters as before except for $c(x) \equiv x^0$) and uniform selection probability, in exploring the potential energy surface in comparison with standard one-way and two-way shooting, using Gaussian selection probability in both cases. In Fig.~\ref{fig:2d_models}F, we show the autocorrelation of the transition channel obtained by computing $\Braket{x^1}$ for each sampled path and assigning the value $+1$ to the path if $\Braket{x^1} > 0$ and $-1$ otherwise. 
The autocorrelation of this observable as a function of the estimated number of force evaluations $\tilde{n}_{\text{F}}$ (multiplying the number of trials by the average number of force evaluations per trial---see Tab.~\ref{tab:force_eval}) is shown in the main panel of Fig.~\ref{fig:2d_models}F, while the inset shows the same quantity as a function of the number of TPS trials $\tau$. The figure shows that, in terms of force evaluations, the algorithms introduced decorrelate the two channels faster than standard two-way and one-way algorithms, with both selection probabilities analyzed. In particular, the use of AAA-TPS with a Gaussian selection probability proves to be the most efficient choice in terms of path-space exploration, both with respect to the total number of force evaluations and the number of TPS trials required. It is worth noting that this improved efficiency is achieved despite a higher average number of force evaluations per trial (reported in Tab.~\ref{tab:force_eval} for the different methods investigated), arising from the longer trajectories generated when shooting from the top of the barrier.
\begin{table}[hbtp]
\caption{\label{tab:force_eval}%
Average number of force evaluations per trials over 50 replicas with the methods investigated.}
\begin{ruledtabular}
\begin{tabular}{cc}
\multirow{2}{*}{TPS Method} & Average number \\
& of force evaluations per trial \\
\hline
Two-Way Shooting\footnotemark[1]& $1424\pm83.8$ \\
One-Way Shooting\footnotemark[1]& $712.8\pm58.3$ \\
Always Reactive Algorithm\footnotemark[1] & $712.7\pm58.3$ \\
Always Reactive Algorithm\footnotemark[2] & $439.0\pm55.3$ \\
Always Accepting Algorithm\footnotemark[1] & $722.5\pm58.2$ \\
Always Accepting Algorithm\footnotemark[2] & $449.0\pm55.6$ \\
\end{tabular}
\end{ruledtabular}
\footnotetext[1]{Gaussian shooting point selection probability.}
\footnotetext[2]{Uniform shooting point selection probability.}
\end{table}
We can now compute $r_{\text{acc}}$ and the ESS for the two selection probabilities and the two algorithms. In addition, for this particular observable we can also compute the integral of the autocorrelation function to take into account correlation effects and obtain an estimate of how many trials and force evaluations are necessary to produce one independent path during the simulation. Results are reported in Tab.~\ref{tab:bsdw_comp_ARAvsAAA}.
\begin{table}[hbtp]
\caption{\label{tab:bsdw_comp_ARAvsAAA}
Acceptance rate $r_{\text{acc}}$ and Effective Sample Size (ESS) normalized by the number of samples $N$ for the methods investigated. Simulations performed on a bi-stable double well (Fig.~\ref{fig:2d_models}D). The integral $\tau_{\text{ch}}$ of the normalized autocorrelation function of the transition channels (Fig.~\ref{fig:2d_models}F) is also reported for the different methods. This provides the number of trials $\tau^{\text{ch}}$ and force evaluations $\tilde{n}^{\text{ch}}_{\text{F}}$ to produce an independent sample.}
\begin{ruledtabular}
\begin{tabular}{ccccc}
\multirow{2}{*}{TPS Method} & \multirow{2}{*}{$r_{\text{acc}}$} & \multirow{2}{*}{$\text{ESS}/N$} & \multirow{2}{*}{$\tau^{\text{ch}}$} & \multirow{2}{*}{$\tilde{n}_{\text{F}}^{\text{ch}}$ $(\cdot 10^5)$}\\ \\
\hline
Two-Way Shooting\footnotemark[1]& $0.28$& $-$ &$216$& $3.08$\\
One-Way Shooting\footnotemark[1]& $0.37$& $-$ &$397$& $2.83$\\
Always-Reactive Algorithm\footnotemark[1]& $0.74$& $-$ &$198$& $1.41$\\
Always-Reactive Algorithm\footnotemark[2]& $0.89$& $-$ &$370$& $1.62$\\
Always-Accepting Algorithm\footnotemark[1]& $-$& $0.61$ &$129$& $0.933$\\
Always-Accepting Algorithm\footnotemark[2]& $-$& $0.85$ &$317$& $1.42$\\
\end{tabular}
\end{ruledtabular}
\footnotetext[1]{Gaussian shooting point selection probability.}
\footnotetext[2]{Uniform shooting point selection probability.}
\end{table}
With both selection probabilities investigated, the methods introduced perform better than state-of-the-art alternatives, drastically reducing the number of force evaluations necessary to produce an independent sample of the transition path ensemble (see last two columns of Table~\ref{tab:bsdw_comp_ARAvsAAA}). An additional reduction of these indicators is observed when substituting the acceptance-rejection step of ARA-TPS with the a posteriori reweighting in AAA-TPS.
In one-way-shooting–based algorithms, the successive replacement of limited path segments causes consecutively sampled trajectories to share a variable, and potentially non-negligible, number of configurations. In Fig.~\ref{fig:2d_models}G, we therefore report the average fractional overlap between a reference path and subsequently sampled trajectories as a function of the number of TPS trials and force evaluations, for standard one-way shooting as well as ARA-TPS and AAA-TPS. Fig.~\ref{fig:2d_models}G shows once again that, in comparison with standard one-way shooting, the introduced algorithms require less force evaluations to obtain a completely new path with respect to a reference one, with both selection probabilities. Since Figure~\ref{fig:2d_models}G represents the \emph{average} overlap after $\tilde{n}_F$ force evaluation, the value where the curves disappear represents the number of force evaluation (trials) necessary to completely replace all the paths in the ensemble using the specific algorithms. Another interesting metric to quantify the efficiency of the methods in path replacement is the average number of force evaluations (trials) necessary to completely replace a path. This is represented by the dotted lines in the main plot (inset) of Fig.~\ref{fig:2d_models}G for the methods investigated. Once again, AAA-TPS with Gaussian selection probability proves to be the most efficient method. 
The shaded regions around the curves in Fig.~\ref{fig:2d_models}G represent the interval between the tenth and ninetieth percentiles of the path-overlap distribution. This visualization highlights the differences between acceptance-rejection–based methods and a posteriori reweighting, particularly for a Gaussian shooting-point distribution. Specifically, the red shaded area (quantifying the width of the overlap distribution over five million path trials obtained with AAA-TPS) is approximately half that obtained with ARA-TPS, which itself is about half of that obtained with standard one-way shooting. While the difference between ARA-TPS and standard one-way shooting is expected given the two-fold increase in the acceptance rate derived by the always-reactive algorithm, the difference between ARA-TPS and AAA-TPS is more noteworthy as it arises directly from the more efficient exploration of path space provided by always accepting proposed trajectories, helping the system escaping faster from low-energy trajectories.

\subsection{Carbon dioxide clathrate hydrates}
\begin{figure*}[htbp]\includegraphics[width=\textwidth]{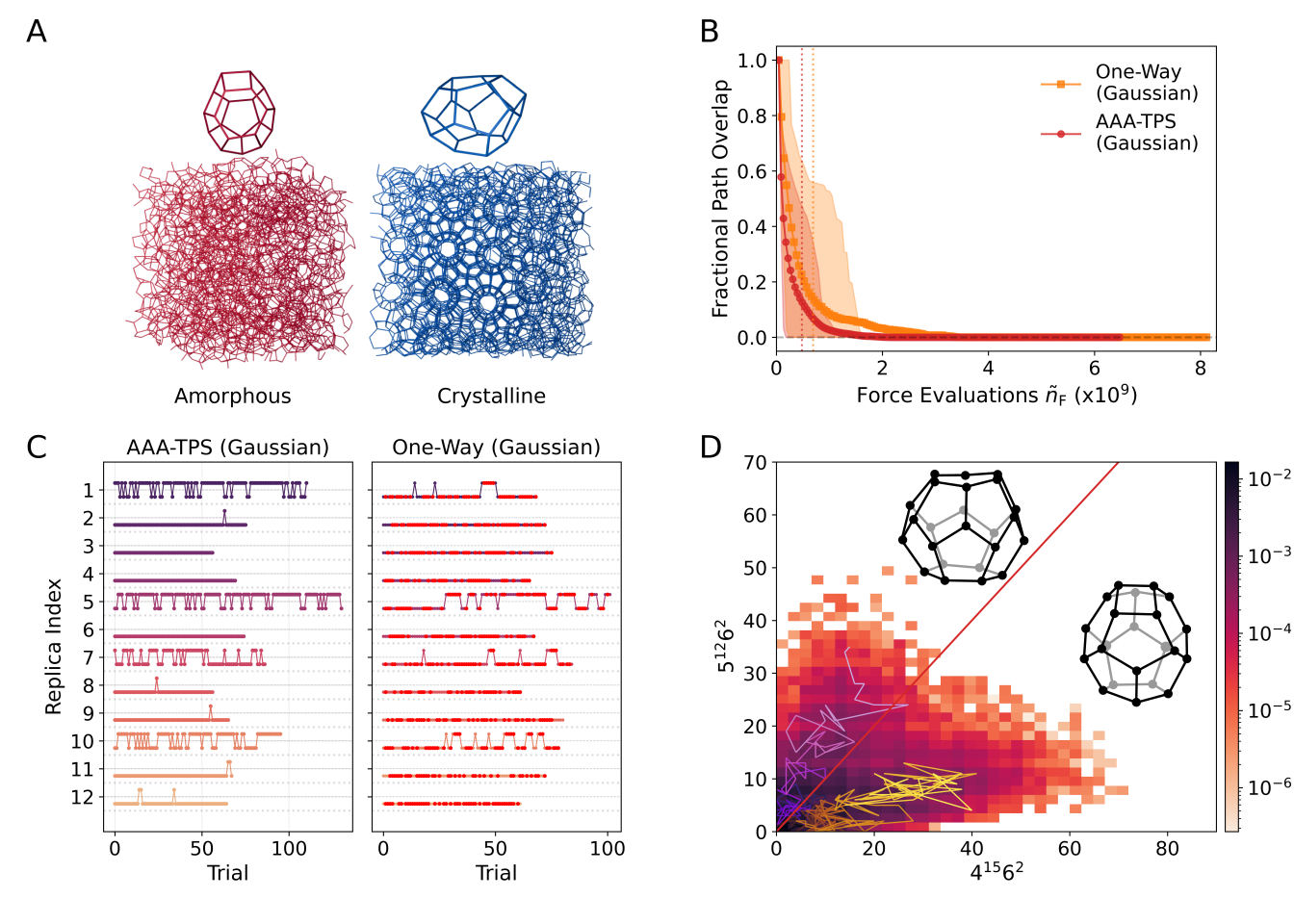}
\caption{\label{fig:Clathrates}Summary of numerical simulation results for the clathrate system. 
(A) Schematic representation of CO$_2$ clathrate hydrate structures, showing amorphous (left) and crystalline (right) phases. This representation shows the arrangement of water cages, with CO$_2$ guest molecules omitted to reveal the host architecture. 
(B) Fractional overlap between a reference path and the successive element of the ensemble as a function of the estimated number of force evaluations in application to a CO$_2$ clathrate hydrate system. Shaded areas represent the 10th and 90th percentiles of the distribution. The vertical dotted lines indicate the average amount of force evaluations needed to fully replace an initial path. 
(C) Transitions between crystalline (upper level) and amorphous (lower level) channels for all replicas. Each point represents an accepted trial within a replica, with the channel state being determined at the trajectory endpoint. Results are shown for the Always Accepting Algorithm (AAA-TPS) (left), and standard one-way shooting algorithm (right). For the one-way shooting algorithm, rejected trajectories are indicated by a red marker.
(D) Histogram of the cage count densities of the $5^{12}6^2$ cage (left illustration) and $4^15^{10}6^2$ cage (right illustration) for  the whole transition paths for the Always Accepting Algorithm (AAA-TPS). The red line indicates an equal number of both cage types. Two distinct trajectories are highlighted, with progression in time indicated by a color gradient from dark to light.}
\end{figure*}

As a further study, we use AAA-TPS to investigate the homogeneous nucleation of CO$_2$ Clathrate Hydrates, crystalline water structures that trap CO$_2$ molecules within hydrogen-bonded cages under high-pressure and low-temperature conditions (Fig.~\ref{fig:Clathrates}A). This system possesses important features for technological
applications, in particular in the context of the current climate crisis~\cite{english_perspectives_2015,sinehbaghizadeh_comprehensive_2023,sloan_fundamental_2003,takeya_preservation_2016}. The study of the nucleation process that forms these kinds of compounds has been the subject of many numerical studies~\cite{lauricella_methane_2014,lauricella_clathrate_2015,peters_path_2008,barnes_reaction_2014}, including different examples using state-of-the-art TPS algorithms~\cite{arjun_molecular_2021,arjun_rate_2020,arjun_unbiased_2019} and it provides a good example to test the proposed methods.
The simulation set up is the same as in Ref.~\citenum{arjun_molecular_2021} and details are reported in the Supplementary Material.

Although this simulation scheme employs underdamped Langevin dynamics, velocity autocorrelations decay completely within \SI{300}{\femto \second}. This confirms that the system investigated behaves fully overdamped on relevant nucleation timescales (hundreds of picoseconds), validating the use of the algorithm proposed. Velocity and position autocorrelation functions are reported in the Supplementary Material.

As in Ref.~\citenum{arjun_molecular_2021}, liquid and solid phases are identified using the MGC-1 collective variable~\cite{barnes_two-component_2014}. Briefly, this quantity measures the local similarity of a water molecule’s hydrogen-bond network to that of structure I gas hydrates by counting nearest-neighbor arrangements consistent with hydrate cages. Further details on the calculation of this order parameter are provided in the Supplementary Material. In particular, MGC-1 values below 10 classify the system as liquid, whereas values above 300 indicate a solid phase.

Initial paths are generated by equilibrating the system at \SI{300}{\kelvin} and \SI{400}{\bar} (promoting a CO$_2$ bubble formation) followed by \SI{250}{\kelvin} and \SI{400}{\bar}. Long MD simulations ($\sim$\SI{1}{\micro \second}) at these values of temperature and pressure reveal spontaneous clathrate hydrate nucleation. To obtain initial paths for TPS at the target conditions (\SI{260}{K} and \SI{500}{\bar}), configurations with clathrate clusters of various sizes from the \SI{250}{\kelvin} trajectory are selected. From each of these, multiple simulations are initiated; trajectory segments exhibiting transitions to both liquid and solid states are then stitched together to form complete liquid-to-solid pathways at the target temperature and pressure. From these initial paths, 12 standard one-way shooting TPS simulations~\cite{bolhuis_transition-path_2003} are started, running for 96 hours on NVIDIA A100 GPUs~\cite{turisini_leonardo_2024}. 

From these 12 equilibrated initial paths, we run 12 independent replicas of TPS simulations for both the AAA-TPS and one-way shooting algorithms with generalized normal (GN) selection probabilities $\mathcal{G}\mathcal{N}(c(x), c_{\text{ref}}, \alpha, \beta)$ with $c(x) \equiv \text{MCG-}1$ and parameters $c_{\text{ref}}=175$, $\alpha = 15$ and $\beta = 2.5$ (details in the Supplementary Material). On the hardware specified above, the one-way shooting method produced \SI{0.385}{\micro \second} of usable trajectory data per day, while the AAA-TPS algorithm generated \SI{1.069}{\micro \second} per day.
This simulation scheme leaves us with 376 accepted and 539 rejected trajectories for one-way shooting, while AAA-TPS produced 973 trajectories, which are always accepted. 

We use these trajectories to compare the path segment replacement efficiency of AAA-TPS and standard one-way shooting for the clathrate formation (Fig. \ref{fig:Clathrates}B). AAA-TPS demonstrates higher efficiency in path replacement, requiring an average of $0.48\cdot10^9$ force evaluations compared to $0.69\cdot10^9$ for one-way shooting TPS to fully replace an original path.

Two distinct nucleation pathways in clathrate hydrates are known to emerge from differing cage population dynamics, namely the crystalline and the amorphous channel. Following previous work~\cite{arjun_molecular_2021}, we distinguish the nucleation channels by comparing the relative populations of two characteristic cages: a higher abundance of the $4^15^{10}6^2$ cage (Fig.~\ref{fig:Clathrates}A, left) indicates the amorphous channel, whereas a higher abundance of the $5^{12}6^2$ cage (Fig.~\ref{fig:Clathrates}A, right) indicates the crystalline channel. Each trajectory obtained from TPS sampling is classified into a channel by counting the number of these cages in its final configuration, using a modified implementation of the GRADE code~\cite{mahmoudinobar_grade_2019}. Through this classification, we can quantify how often transitions between channels occur from one trial to the next (see Fig.~\ref{fig:Clathrates}C). The numbering of the TPS runs links the two methods to their corresponding initial conditions, such that, for example, batch 1 for both methods originates from the same initial path. We observe that batches generally fall into two categories: those with a high likelihood of channel switching and those with a low likelihood, with many complete TPS runs exhibiting zero channel switches. The similar behavior observed across batches in both methods suggests that the switching dynamics is correlated with the choice of initial path. We therefore classify batches 1, 5, 7, and 10 as belonging to the high-switching group, while all remaining batches are assigned to the low-switching group. We define the switching probability as the ratio of observed channel switches divided by the total number of trial-to-trial transitions within a batch. This analysis reveals that the low-switching group exhibits a mean switching probability of $2.3 \% \pm 2.1 \%$ for AAA-TPS and
no switches for one-way shooting. In the high switching group the corresponding values are $33.2 \% \pm 1.9 \%$ for AAA-TPS, compared to only $12.9 \% \pm 3.4 \%$ for one-way shooting. 

This greatly increased switching probability leads to a greatly increased sampling of the crystalline channel. As shown in Fig.~\ref{fig:Clathrates}D, this is also evident in the histogram of the two relevant cage types observed along all trajectories, where cage populations are recorded every \SI{2}{\nano \second}. The vertical axis shows the number of $5^{12}6^2$ cages,
characteristic for the crystalline channel, while the horizontal axis shows the number of $4^15^{10}6^2$ cages,
characteristic for the amorphous phase. The red line indicates the channel divide, such that the crystalline channel lies above the line and the amorphous channel below. Comparing with the results reported in the upper left panel of Fig.~2 in Ref.~\citenum{arjun_molecular_2021} we see how AAA-TPS is able to explore regions of the transition path ensemble which were previously hard to investigate with the available techniques. In the Supplementary Material we also provide the cage-type histogram obtained with standard one-way shooting algorithm obtained from the simulation performed in this work, fully compatible with Ref.~\citenum{arjun_molecular_2021}. The increased exploration of the crystalline channel in CO$_2$ hydrate formation at \SI{260}{\kelvin} and \SI{500}{\bar} is enabled by the higher efficiency of path sampling provided by AAA-TPS, which makes accessible pathways that were nearly unreachable with previously employed standard TPS techniques.

Finally, to estimate the efficiency of employing the reweighting procedure for this systems, rather than relying solely on the acceptance-rejection procedure of ARA-TPS, we again compute the effective sample size (ESS) as defined in Eq.~\ref{eq:ess}. This analysis yields an ESS/$N$ value of 0.67. Assuming a doubling of the acceptance ratio observed for the one-way shooting algorithm, we estimate that ARA-TPS would achieve an acceptance rate of $r_\text{acc} = 0.82$. However, as demonstrated by the results for the two-dimensional model system, such metrics alone are insufficient to reliably predict the performance of ARA-TPS for CO$_2$ clathrates. Instead, direct simulations are required to explicitly assess correlation effects before any definitive conclusions can be drawn.

\section{Discussion}
\label{sec:disc}
The numerical results obtained here for different models indicate that the algorithms presented here are 2-4 times more efficient than state-of-the-art alternatives. The improved efficiency arises from two factors: ARA-TPS takes advantage of overdamped dynamics to relax the need of choosing a shooting direction before the path generation, resulting in a two-fold increase in the number of accepted paths; AAA-TPS, on the other hand, introduces an \emph{a posteriori} reweighting procedure that removes waste in trajectory production. Standard TPS methods perform the two main steps of Monte Carlo in one go: they produce candidates for path space exploration and immediately assign to them the correct statistical weights by an acceptance-rejection step. In practice, the correct statistical weight is assigned to the current configuration by proposing a new configuration and comparing their relative probabilities. 
To avoid the waste of resources related to the rejection of trajectories, AAA-TPS detaches the exploration step and the weight assignment: It first explores path space by generating samples of the transition path ensemble that are always accepted. Then it assigns to each of the samples produced the correct statistical weight, which can be computed based only on properties of the specific trajectory. The procedure proposed here differs fundamentally from other waste-recycling Monte Carlo approaches, most notably Ref.~\citenum{frenkel_waste-recycling_2006}, in which information from rejected moves is incorporated into estimators, while the exploration of state space itself does not benefit from these moves, as they are still explicitly rejected. It also differs from other rejection-free path-ensemble sampling strategies~\cite{lazzeri_optimal_2025}, where the explored space is not restricted to transition paths but includes all trajectories, even non-reactive ones.

One might reasonably wonder, at this point, if this does not completely overturn the whole principle behind Monte Carlo algorithms, namely the idea of importance sampling. Indeed, rejecting moves keeps the Markov Chain from exploring highly unlikely samples and focuses computational resources on the most important portion of the target space. Although this question is legitimate, this does not take into account the cornerstone of the TPS algorithm compared to standard configuration Monte Carlo methods,\footnote{An exception in this sense is represented by Hamiltonian Monte Carlo~\cite{duane_hybrid_1987}, where proposal moves are obtained by evolving the system along trajectories generated by Hamiltonian dynamics, in order to move the system along isoprobability hyper-surfaces.} namely the use of the natural equilibrium dynamics to propose new samples. Every trajectory proposed is generated using the true underlying dynamics of the system, and therefore the probability of generating highly unlikely moves is as high as the dynamics allows. In combination with the use of an unperturbed shooting point, this guarantees the stationarity of the sampled distribution, even without acceptance-rejection schemes. An additional result of this procedure is that explored paths, however low their weight might be in the final ensemble, correspond to possible reaction channels for the transition, and are thus worth exploring. 

However, one should not expect the reweighting procedure to always be more efficient than an acceptance-rejection approach. It is easy to imagine cases where the distributions $\widetilde{P}_{\mathrm{AB}}[X(L)]$ and $P_{\mathrm{AB}}[X(L)]$ are so different that a reweighting procedure is mostly ineffective (or, in other words, the ESS approaches zero). A qualitative argument can be provided by analyzing the ESS in case of a uniform selection probability. It can be shown that Eq.~\eqref{eq:ess} for the ESS can be recast in terms of the averages of the weights over the sampled distribution $\widetilde{P}_{\mathrm{AB}}[X]$. Expressing numerator and denominator by arithmetic averages over the number of samples $N$ obtained sampling from the distribution $\widetilde{P}_{\mathrm{AB}}[X]$, we get
\begin{equation}
\label{eq:ess_ptilde}
\text{ESS} = N \frac{\Braket{\Omega\inv}^2_{\widetilde{P}_{\mathrm{AB}}}}{\Braket{\left(\Omega\inv\right)^2}_{\widetilde{P}_{\mathrm{AB}}}},
\end{equation}
where $\braket{\cdot}_{\widetilde{P}_{\mathrm{AB}}}$ denotes the average over $\widetilde{P}_{\mathrm{AB}}[X]$. These averages can be written in terms of averages over the target distribution $P_{\mathrm{AB}}[X]$ by enforcing the definition of $\widetilde{P}_{\mathrm{AB}}[X] = \Omega[X]P_{\mathrm{AB}}[X]$. We first have
\begin{equation}
\begin{split}
\Braket{\Omega^{-1}}_{\widetilde{P}_{\mathrm{AB}}} &= \sum_L\int\dd x_0\dots\dd x_{L-1} \Omega\inv[X] \widetilde{P}_{\mathrm{AB}}[X] \\
&= \sum_L\int\dd x_0\dots\dd x_{L-1} P_{\mathrm{AB}}[X] =1,
\end{split}
\end{equation}
since the distribution $P_{\mathrm{AB}}[X]$ is normalized. Then
\begin{equation}
\begin{split}
\Braket{\Omega^{-2}}_{\widetilde{P}_{\mathrm{AB}}} &= \sum_L\int\dd x_0\dots\dd x_{L-1} \left(\Omega\inv[X]\right)^2 \widetilde{P}_{\mathrm{AB}}[X] \\
& \!\!\!\!\!\!\!\!\!\!\!\!\!\!\!\!\!= \sum_L\int\dd x_0\dots\dd x_{L-1} \Omega\inv[X] P_{\mathrm{AB}}[X] = \Braket{\Omega\inv},
\end{split}
\end{equation}
where $\braket{\cdot}$ denotes the mean value over the target distribution $P_{\mathrm{AB}}[X]$. The ESS can then be written as
\begin{equation}
\label{eq:ress_uniform}
\mathrm{ESS} = \frac{N}{\Braket{\Omega\inv}} = N \frac{1}{\Braket{L}}\Braket{\frac{1}{L(X)}}\inv,
\end{equation}
where in the second equality we have assumed a uniform selection probability, namely $\Omega[X(L)] = L(X)$, and we have normalized the weights by $\Braket{L}$ to make them dimensionless.
We can now recast the denominator of Eq.~\eqref{eq:ress_uniform} in the form 
\begin{equation}
\Braket{L}\Braket{\frac{1}{L(X)}} = \Braket{L}\Braket{\frac{1}{\Braket{L} + \Delta L(X)}} = \Braket{\frac{1}{1+\frac{\Delta L(X)}{\Braket{L}}}},
\end{equation}
which, for small $|\Delta L(X)|^n/\braket{L}^n$ for $n\ge2$ (namely, fluctuations around the mean are small for all moments retained), can be Taylor-expanded as
\begin{equation}
\frac{\text{ESS}}{N} = 1 - \frac{\Braket{\Delta L(X)^2}}{\Braket{L}^2} + \frac{\Braket{\Delta L(X)^3}}{\Braket{L}^3} + \mathcal{O}(\Braket{\Delta L(X)^4}),
\end{equation}
which shows that the effective sample size decreases for large relative variance and for negative relative skewness (i.e. the path length distribution asymmetrically shifted to long paths compared to the average path length), while a small relative variance and a positive skewness have the effect of maximizing it. This is not surprising considering the acceptance criterion of standard acceptance-rejection TPS methods with uniform shooting point selection probability $\min[1, L/L']$, which keeps the Markov chain from exploring path with very long transition times (see also Supplementary Material). In this context, for very broad or negatively skewed path-length distributions associated with $\widetilde{P}_{\mathrm{AB}}[X(L)]$, assessing the efficiency of \emph{a posteriori} reweighting relative to acceptance-rejection schemes should also account for the additional computational cost arising from the generation of longer trajectories. This effect is naturally captured when efficiency is evaluated in terms of force evaluations, a metric that should anyway always be preferred over the commonly used number of TPS trials.

\section{Conclusion}
In this work, we have introduced two transition path sampling algorithms that enhance the efficiency of path sampling for arbitrary shooting-point probability distributions, and we have discussed possible metrics for assessing their efficiency. From an implementation perspective, both methods are particularly simple to incorporate into existing TPS codes. The modified path generation scheme of ARA-TPS requires only a dynamic selection of which part of the old trajectory must be stitched to the newly generated segment, based on which stable state has been reached first by the new trajectory. In contrast, the a posteriori reweighting approach only involves a trivial modification (namely, the removal of the acceptance criterion) together with the storage of the normalization factors associated with the shooting-point selection probability (namely, the denominator of the RHS of Eq.~\eqref{eq:psel_w}), a quantity that is already computed in current TPS implementations.
As a result, both methods require only minimal changes relative to standard one-way shooting schemes, as their algorithmic backbone remains very close to conventional TPS workflows. This makes them not only efficient but also highly accessible from an implementation standpoint.

By combining the proposed efficiency metrics with the simplicity of the algorithms, one can envision adaptive TPS schemes in which simulations are initially performed using ARA-TPS, correlations of the observables under investigation are monitored on the fly, and the corresponding AAA-TPS weights are estimated from Eq.~\eqref{eq:weights_ratio_final}. This procedure further enables an estimate of the effective sample size that would be obtained under AAA-TPS using the general expression of Eq.~\eqref{eq:ress_uniform}, namely $\text{ESS}/N = \braket{\Omega\inv}\inv$. Such an approach naturally lends itself to an automated switching strategy, whereby the simulation dynamically transitions between ARA-TPS and AAA-TPS once it becomes evident that one method provides a significantly higher sampling efficiency than the other.

As demonstrated for carbon dioxide clathrate hydrates, we here emphasize once more that the requirement of an overdamped regime for the use of the algorithms introduced in this work does not apply to the underlying system dynamics, but rather to the specific transition process under investigation. ARA-TPS (with or without reweighting) can be used whenever the rare event of interest happens over time scales larger than the typical decay of the velocity autocorrelation function. Events where inertial effects are important, such as catalytic reactions~\cite{basner_how_2005}, dissociation of molecules~\cite{geissler_autoionization_2001}, proton transfer~\cite{dellago_kinetics_2006}, etc., must be approached with standard algorithms, as the continuity of velocities between the two segments of the paths (e.g., $X_{\text{bw}}$ and $X'_{\text{fw}}$) becomes crucial in the generation step to obtain physically meaningful trajectories. Nonetheless, in these cases, always accepting the reactive paths generated and reweighting a posteriori could still represent an advantageous choice by speeding up the exploration of path space and helping decorrelating quickly reactive trajectories. However, rare events happening on time scales larger than the standard time of decay of velocity autocorrelation are very common in physics and extremely relevant for technological application. In addition to the hydrates formation discussed in this work and other similar homogeneous and heterogeneous nucleation phenomena~\cite{diaz_leines_template-induced_2022}, other examples are biomolecular reorganization~\cite{quaytman_reaction_2007, best_coordinate-dependent_2010}, phase transitions in ferromagnetic spin systems~\cite{mora_transition_2012,desplat_path_2020,moritz_interplay_2017}, ion transport through membrane channels~\cite{marti_transition_2004} and many more.

\section*{DATA AVAILABILITY}
The data that support the findings of this study together with the source code used for the numerical simulations are available via the project \href{https://github.com/CompPhysVienna/paper_AAA-TPS}{GitHub page}.

\begin{acknowledgments}
We acknowledge access to LEONARDO at CINECA, Italy, via AURELEO (Austrian Users at LEONARDO supercomputer) projects n. 72451 and n. 72829. Financial support of the Austrian Science Fund (FWF) through the SFB TACO, Grant No. 10.55776/F8100 is also gratefully acknowledged.
\end{acknowledgments}

\newpage

\onecolumngrid
\normalsize
\patchcmd{\large}{15}{15}{}{}
\begin{center}
  \textbf{\LARGE Supplementary Material: An Always-Accepting Algorithm for Transition Path Sampling}\\[.2cm]
  Magdalena H\"aupl$^{1,2}$, Sebastian Falkner$^{1,2}$, Peter G. Bolhuis$^{3}$, Christoph Dellago$^{1,4}$, and Alessandro Coretti$^{1,*}$\\[.1cm]
  {\itshape ${}^1$University of Vienna, Faculty of Physics, 1090 Vienna, Austria.\\
  \itshape ${}^2$Institute of Physics, University of Augsburg, Universit\"atsstraße 1, 86159 Augsburg, Germany.\\
  \itshape ${}^3$Van ’t Hoff Institute for Molecular Sciences, University of Amsterdam, PO Box 94157, 1090GD Amsterdam, The Netherlands.\\
  \itshape ${}^4$Research Platform on Accelerating Photoreaction Discovery (ViRAPID), University of Vienna, 1090 Vienna, Austria

  }
  ${}^*$Electronic address: alessandro.coretti@univie.ac.at\\
(Dated: \today)\\[2cm]
\end{center}

\def\section{\@startsection
  {section}{1}{\z@}%
  {-3.5ex plus -1ex minus -.2ex}
  {2.3ex plus .2ex}
  {\normalfont\Large\bfseries\raggedright}}

\def\subsection{\@startsection
  {subsection}{2}{\z@}%
  {-3.25ex plus -1ex minus -.2ex}%
  {1.5ex plus .2ex}%
  {\normalfont\normalsize\bfseries}}

\def\subsubsection{\@startsection
  {subsubsection}{3}{\z@}%
  {-3.25ex plus -1ex minus -.2ex}%
  {1.5ex plus .2ex}%
  {\normalfont\normalsize\itshape}}

\makeatother

\setcounter{equation}{0}
\setcounter{figure}{0}
\setcounter{table}{0}
\setcounter{page}{1}
\setcounter{section}{0}
\renewcommand{\theequation}{S\arabic{equation}}
\renewcommand{\thefigure}{S\arabic{figure}}
\renewcommand{\thetable}{S\arabic{table}}
\renewcommand{\bibnumfmt}[1]{[S#1]}
\renewcommand{\citenumfont}[1]{S#1}
\renewcommand{\thesection}{S\Roman{section}}
\renewcommand{\thepage}{S\arabic{page}}

\titleformat*{\section}{\Large\bfseries}

\setlength{\parindent}{0cm}
\setlength{\parskip}{8pt}

\section{Simulation Details}
\subsection{Two-dimensional Models}
All simulations for the 2D models are run using the Euler–Maruyama integrator for overdamped Langevin dynamics~\cite{sm_leimkuhler_molecular_2015} 
\begin{equation}
X_{n+1} = X_{n} + \frac{F(X_n)}{\gamma}\Delta t + \xi_n\sqrt{\frac{2k_BT}{\gamma}\Delta t} = X_n + \beta D F(X_n) \Delta t + \xi_n\sqrt{2D\Delta t},
\end{equation}
where $\xi_n$ is white noise, $D = \frac{k_BT}{\gamma}$ and $\beta = (k_BT)\inv$. The simulations on the two-dimensional models are performed in reduced units and we then set $D=1$, $\beta = 1$ and $\Delta t = 0.001$. Stable states are identified by $\zeta^2 - R^2 < 0$ where $\zeta^2$ is given by the expression
\begin{equation}
\label{eq:states}
\zeta^2(x^{(0)}, x^{(1)}; a, \theta, c) = \left(\frac{\hat{x}^{(0)}(\theta, c)}{a^{(0)}}\right)^2 + \left(\frac{\hat{x}^{(1)}(\theta, c)}{a^{(1)}}\right)^2,
\end{equation}
where
\begin{equation}
\begin{pmatrix}
\hat{x}^{(0)}(\theta, c) \\
\hat{x}^{(1)}(\theta, c)
\end{pmatrix}
= \hat{x}(\theta, c) = \mathbf{R}(\theta)(x-c) = \left\{
\begin{aligned}
\cos(\theta)(x^{(0)} - c^{(0)}) + \sin(\theta)(x^{(1)}  - c^{(1)})\\
-\sin(\theta)(x^{(0)} - c^{(0)}) + \cos(\theta)(x^{(1)} - c^{(1)})
\end{aligned}
\right.,
\end{equation}
with $a$, $\theta$ and $c$ tunable parameters. 
All TPS simulations are performed by running 50 replicas with $10^5$ trial each equilibrated independently for $2.5 \times 10^3$ trials with the selected algorithm. Reference TPS ensemble for the standard double well is obtained by running 250 replicas with the same number of sampling and equilibration trials as above using two-way shooting TPS with uniform shooting point selection probability. Paths and observables of interest are saved every trial. The maximum path length is set to $2.5\times10^4$ frames for both models and is never hit during TPS simulations.

\subsubsection{Standard Double Well}
The standard double well potential has the form:
\begin{align}
    U(x) = \alpha \bigl\{\bigl(x^{(0)} - x^{(1)}\bigl)^2 + \bigl[(x^{(0)})^2 -1\bigr]^2\bigr\},
\end{align}
where $\alpha$ is adjusted to match the desired barrier height, namely $\alpha=10$. States $\mathrm{A}$ and $\mathrm{B}$ are obtained by Eq.~\eqref{eq:states} setting $R^2=0.05$, $a=(1,2)$, $\theta=-0.25$, and $c_{\mathrm{A}} = (-1,-1)$ and $c_{\mathrm{B}} = (1,1)$.

\subsubsection{Bi-stable Double Well}
The bi-stable double well potential has the form:
\begin{align}
    U(x) = \frac{\alpha}{8} \bigl\{\beta\bigl[(x^{(0)})^2 + (x^{(1)})^2 - 4)^2\bigl] + (x^{(1)})^2\bigr\},
\end{align}
where we set $\alpha=15$ and $\beta=0.25$. States $\mathrm{A}$ and $\mathrm{B}$ are obtained by Eq.~\eqref{eq:states} setting $R^2=0.15$, $a=(1,1)$, $\theta=0$, and $c_{\mathrm{A}} = (-2,0)$ and $c_{\mathrm{B}} = (2,0)$.

\subsection{CO${_2}$ Clathrate Hydrates}
\subsubsection{Simulation Details}

The simulation cell consists of a cubic, periodic box, populated with 2944 water molecules and 512 CO$_2$ molecules. Trajectories are generated using the OpenMM 8.2.0 simulation package~\cite{sm_eastman_openmm_2024}. Water-water interactions are modeled via the TIP4P/Ice force field~\cite{sm_abascal_potential_2005}, while guest-guest interactions employ the TraPPE force field~\cite{sm_potoff_vaporliquid_2001}. Modified Lorentz-Berthelot combining rules ($\chi$ = 1.08) are applied to guest-water interactions~\cite{sm_costandy_role_2015}. The water oxygen-hydrogen and hydrogen-hydrogen distance were constrained using OpenMM at 0.09572 $\si{nm}$ and 0.15139 $\si{nm}$, respectively as prescribed by the TIP4P/Ice force field. The van der Waals cutoff was set to 1 \si{nm} and the electrostatic interactions were calculated using the Particle Mesh Ewald method. Similarly, the CO$_2$ bond length was constrained at 0.116 $\si{nm}$ and the angle between the two bonds was harmonically restrained at 180° with a spring force of 1500 $\si{kJ mol^{-1} rad^{-2}}$, as prescribed by the TraPPE force field. 

To sample the NPT ensemble, the Velocity Verlet with the Velocity Randomization integrator~\cite{sm_sivak_time_2014} is used, acting also as a thermostat, with a timestep of \SI{2}{\femto \second} and a frequency of \SI{1}{\per \pico \second}. A Monte Carlo barostat is employed at a frequency of \SI{0.25}{\per \pico \second} to keep the pressure constant.

\subsubsection{Mutually Coordinated Guest}
As commonly done in other works on hydrates~\cite{sm_barnes_reaction_2014}, we use clusters of Mutually Coordinated Guests (MCG) to identify the stable states for the TPS study of the nucleation of CO${}_2$ clathrate hydrates.
\begin{figure}[tbp]
    \centering

    \scalebox{0.9}{
    \begin{minipage}[b]{0.48\textwidth}
        \centering
        \begin{subfigure}[b]{0.9\linewidth}
            \renewcommand\thesubfigure{a}
            \centering
            \includegraphics[width=\linewidth]{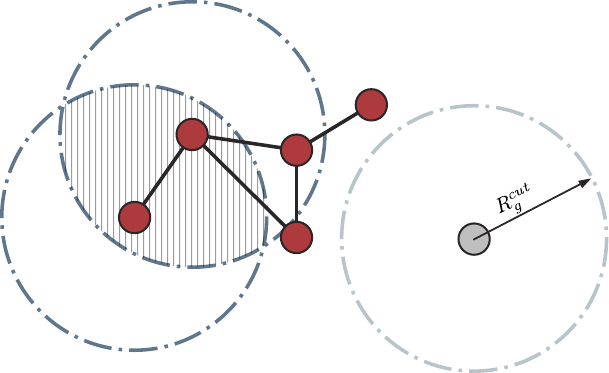}
            \caption{}
            \label{fig:left_top_mcg}
        \end{subfigure}

        \vspace{0.5cm}

        \begin{subfigure}[b]{0.7\linewidth}
            \renewcommand\thesubfigure{c}
            \centering
            \includegraphics[width=\linewidth]{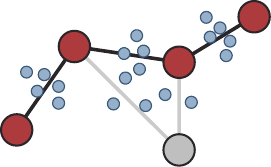}
            \caption{}
            \label{fig:left_bottom_mcg}
        \end{subfigure}
    \end{minipage}
    \hfill
    \begin{minipage}[b]{0.48\textwidth}
        \centering
        \begin{subfigure}[b]{\linewidth}
            \renewcommand\thesubfigure{b}
            \centering
            \includegraphics[width=\linewidth]{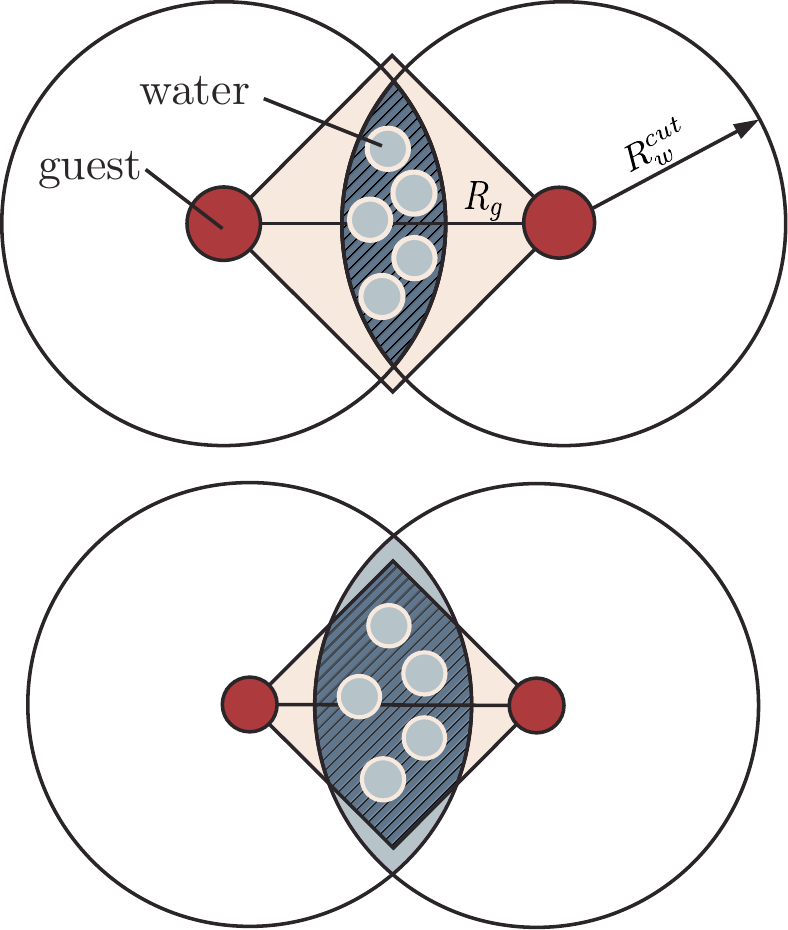}
            \caption{}
            \label{fig:right_mcg}
        \end{subfigure}
    \end{minipage}
    }
    \caption{This visualization outlines the three-step procedure for determining the MCG order parameter, which quantifies the size of the largest cluster. (a) Guest Condition: Guest molecules (spheres) within pairwise distance $R_g^{cut}$ = \SI{9}{\angstrom} (dashed blue circles) form preliminary edges (black lines). (b) Water Condition: Each edge requires at least $N_w = 5$ water molecules simultaneously located within
    $R_w^{cut}$ = \SI{6}{\angstrom} of both guests (transparent spheres)
    and inside overlapping conical volumes spanning $45^{\circ}$ radially outwards from the inter-guest axis (dark blue shading).
    (c) Edges are retained only when sufficient bridging waters exist. Guests with at least $N$ connections ($N=1$ for MCG-1) become MCG monomers. The largest connected cluster (red circles) defines the order parameter.}
    \label{fig:combined_mcg}
\end{figure}
A cluster is defined here as a collection of MCG monomers, which are guest molecules that meet the following criteria:

\begin{enumerate}
    \item All guest molecules which are within cutoff distance ($R_g^{cut} = 9\,\mathrm{\AA}$) from one another are considered to be connected. Every such connection is referred to as an edge. See Fig. \ref{fig:left_top_mcg} for an illustration of this step.
    \item Not all edges are retained. An edge between two guest molecules is only kept if at least $N_w$ water molecules (with $N_w = 5$ in this study) are found that satisfy both of the following conditions: 
    \begin{itemize}
        \item Each water molecule lies within cutoff distance ($R_w^{cut} = 6\,\mathrm{\AA}$) from both guest molecules.
        \item Each water molecule lies within the overlapping volume of two cones projected toward one another along the axis connecting the guest molecules, each spanning $45^\circ$ from that axis.
    \end{itemize}
    The combination of these two conditions form
    a volume, displayed in Fig. \ref{fig:right_mcg} for two different guest distances as the shaded blue region, in which all $N_w$ water molecules separating the guest molecules must be located. If less than $N_w$ waters are found within this region, the edge is removed, as seen in Fig. \ref{fig:left_bottom_mcg}.
    \item Finally, all guest molecules that have at least $N$ connecting nodes, where $N$ is defined by the MCG-$N$ order parameter ($N=1$ for MCG-1 as used in this study), will be denoted as MCG monomers.

\end{enumerate}
The MCG order parameter is now taken to be the size of the largest cluster of MCG monomers found through this procedure in the system.

\subsubsection{Cage Number Difference
\label{sec:cage_ratio}
}

The nucleation landscape of clathrate hydrates features a complex free energy surface governed by competing cage formation pathways. For CO$_2$ clathrates, we know that there exist at least two different channels of nucleation. The system can either solidify into an amorphous solid, or into a the crystalline phase, generally denoted by sI~\cite{sm_sloan_clathrate_2007}. 

Previous research has established that these nucleation channels are defined by their populations of characteristic cages specific to each structure \cite{sm_zhang_microcanonical_2015,sm_sarupria_homogeneous_2012}. For CO$_2$ clathrates, the amorphous pathway is characterized by a cage type rarely prominent in other hydrate systems: the $4^15^{10}6^2$ cage. The reason for this cage to be unusually common for the CO$_2$ guest molecule, is known to be linked to the size and especially the linear shape of CO$_2$ being compatible with the elongated geometry of the cage \cite{sm_he_what_2017, sm_arjun_molecular_2021}.
The crystalline pathway, on the other hand, shows a high abundance of $5^{12}6^2$ cages, which are the most common cages in sI hydrates \cite{sm_cai_instrumental_2022}.

We define the Cage Difference (CD) metric as the numerical difference between $5^{12}6^2$ and $4^15^{10}6^2$ cage populations:
\begin{equation}
    \mathrm{CD} = N_{5^{12}6^2} - N_{4^15^{10}6^2}.
\end{equation}
A CD greater than 0 indicates that the system is currently in the crystalline nucleation pathway, while a CD below 0 indicates that the system is in the amorphous nucleation pathway.
This parameter is comparable to past studies, which have instead used a ratio of cage numbers, as $\mathrm{CR} = N_{5^{12}6^2} / N_{4^15^{10}6^2}$ \cite{sm_arjun_molecular_2021, sm_zhang_microcanonical_2015, sm_sarupria_homogeneous_2012}. We choose to use the CD instead, since this preserves linear symmetry across the classification threshold and it eliminates sensitivity to division-by-zero artifacts and statistical outliers.

Our cage analysis employs a modified version of the GRADE code \cite{sm_mahmoudinobar_grade_2019}, extended to include identification of the $4^15^{10}6^2$ cage type. This code is able to identify cages in the system by employing a set of rules, which are based on the geometric properties of the cages. The code first identifies all water molecules within a cutoff distance of 3.5 $\,\mathrm{\AA}$ from each other and then identifies whether these molecules form a ring (square, pentagon, hexagon). Checking these rings for relevant overlaps with other rings leads to the discovery of certain `cups', which are parts of the cages we want to identify. Specifically, we identify $5^6$, $5^66^1$ and $4^15^{4}6^2$ cups. $5^{12}6^2$ cages consist of two $5^66^1$ cups and $4^15^{10}6^2$ are made up of a combination of a $5^6$ and a $4^15^{4}6^2$ cup. By finding relevant overlaps between these cups, the code is able to identify the cages in the system. 

\begin{figure}[htbp]
    \centering
    \begin{subfigure}{0.4\textwidth} 
        \centering
        \includegraphics[width=\linewidth]{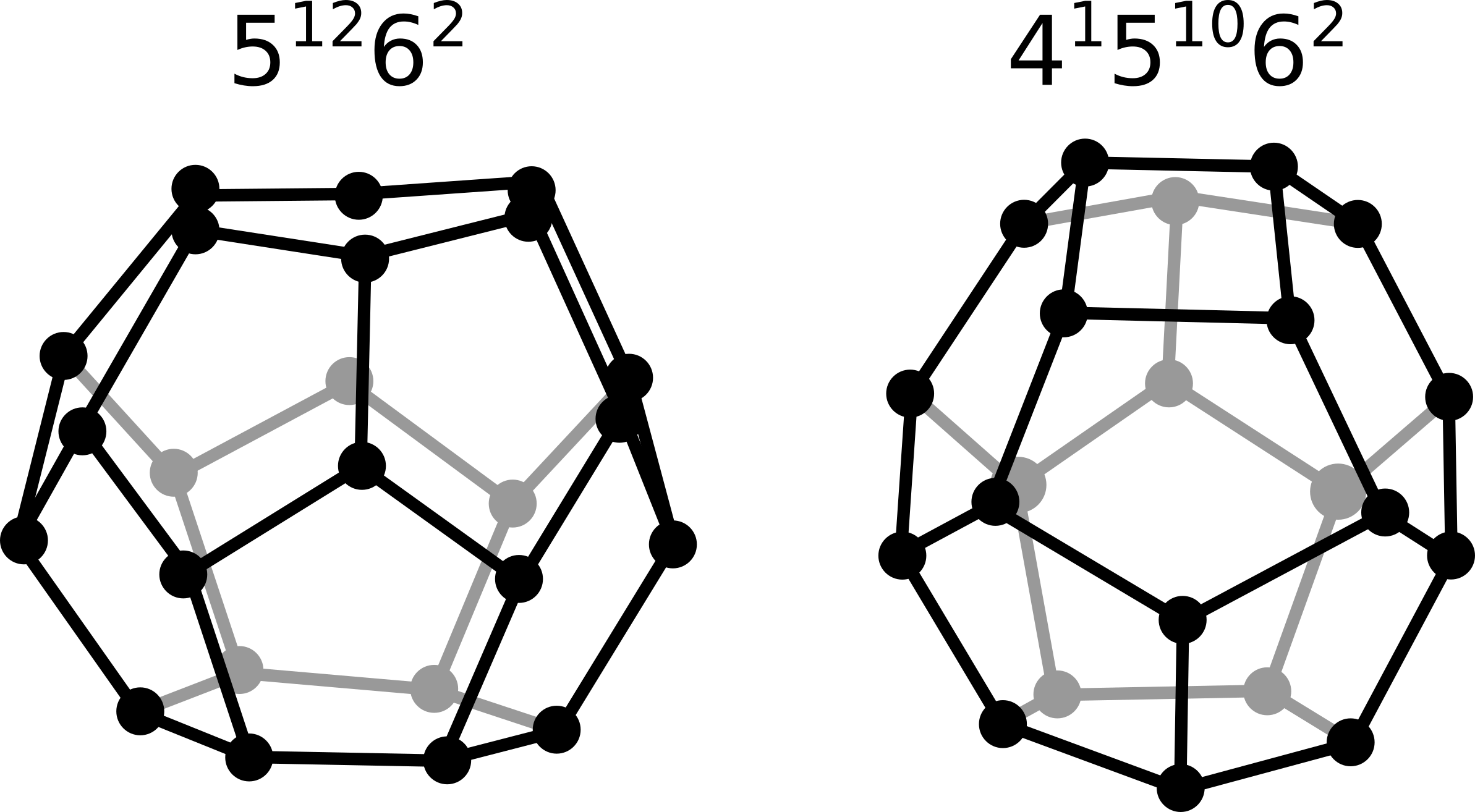}
        \qquad\\
        \qquad \\
        \qquad\\
        \qquad\\
        \caption{}  
        \label{fig:scenario_a}
    \end{subfigure}
    \qquad\qquad
    \begin{subfigure}{0.4\textwidth}
        \centering
        \includegraphics[width=\linewidth]{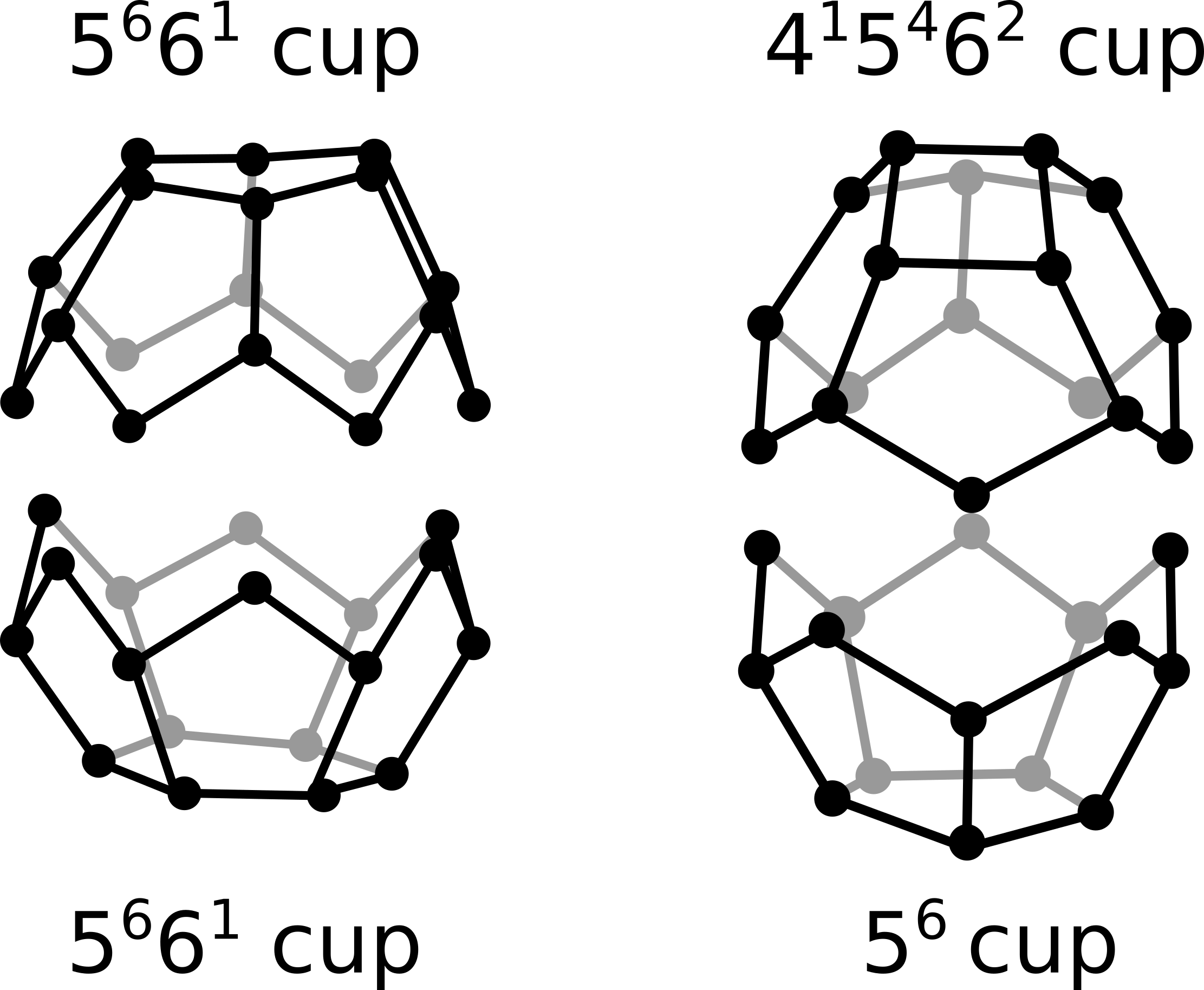}
        \caption{} 
        \label{fig:scenario_b}
    \end{subfigure}

    \caption{(a) Geometric representation of the two most relevant cages in the system, $5^{12}6^2$ and $4^15^{10}6^2$, with nodes representing the water molecules and (b) the cups making up these cages.}
    \label{fig:main_cages}
\end{figure}

\subsubsection{Generalized normal shooting point selection probability}
The generalized normal (GN) distribution offers greater flexibility as a shooting-point selection probability compared to a standard normal distribution. The generalized normal is defined as
\begin{equation}
    \mathcal{G}\mathcal{N}(c(x), c_{\text{ref}}, \alpha, \beta) \propto e^{-(|c(x)-c_{\text{ref}}|/\alpha)^{\beta}},
    \label{eq:generalized_gaussian}
\end{equation}
\begin{figure}[htbp]
    \centering
    \includegraphics[width=0.6\linewidth]{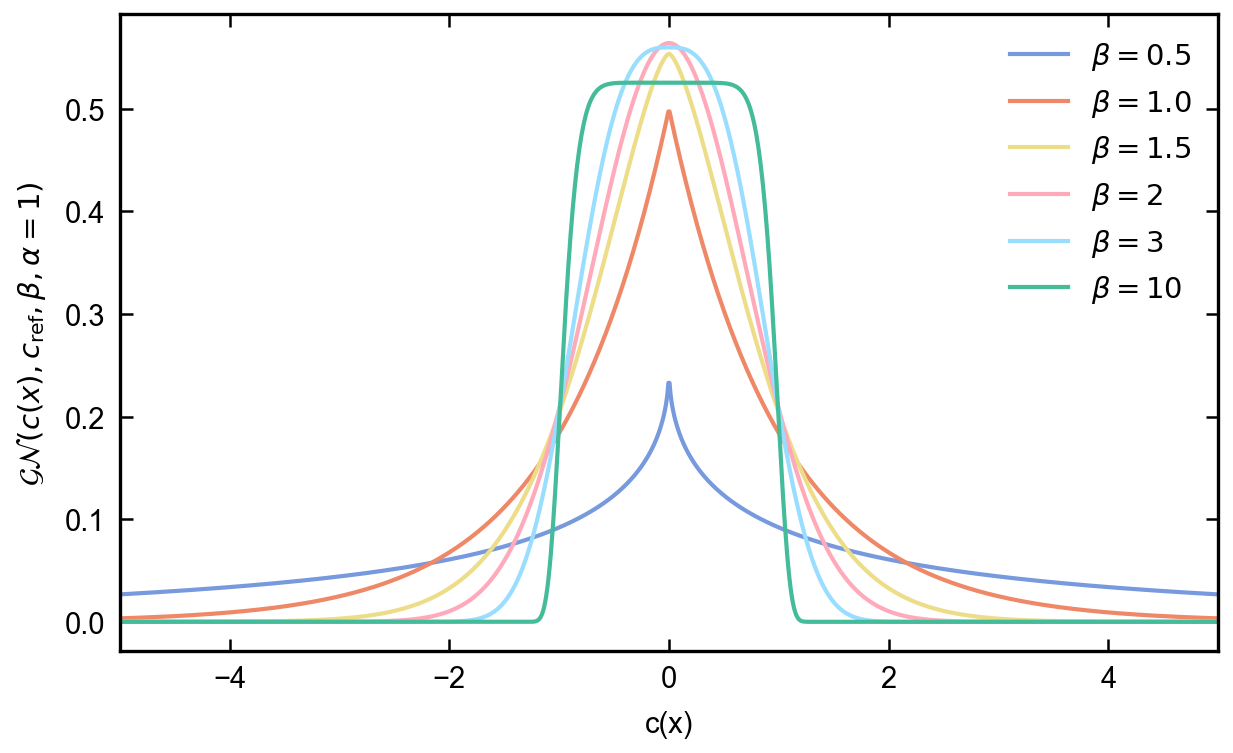}
  \caption{The generalized normal distribution $\mathcal{G}\mathcal{N}(c(x), c_{\text{ref}}, \alpha, \beta)$ for $c_{\text{ref}} = 0$, $\alpha = 1$ and a range of $\beta$ values.}
    \label{fig:gg}
\end{figure}
where $c(x)$ is a collective variable from the space of configurations to $\mathbb{R}$, $c_{\text{ref}} \in 	\mathbb{R}$ denotes the mean, $\alpha \in \mathbb R_{> 0}$ represents the scale, and $\beta \in \mathbb R_{> 0}$ is related to the shape of the function (see Fig.~\ref{fig:gg}).

\section{Additional results}
\subsection{Path Lengths Distribution for Two-dimensional Models}
We show here the path length distribution for the two-dimensional models investigated. We also include the results arising for the always-accepting algorithm before reweighing to show that, in agreement with the theoretical discussion in Sec.~V of the main text, paths characterized by a longer path length are overrepresented. The reweighing procedure accounts for that and the path length distribution obtained with AAA-TPS is perfectly superposed with the ones obtained with the other method, based on acceptance/rejection schemes.

\begin{figure}[htbp]
    \centering
    \includegraphics[width=\linewidth]{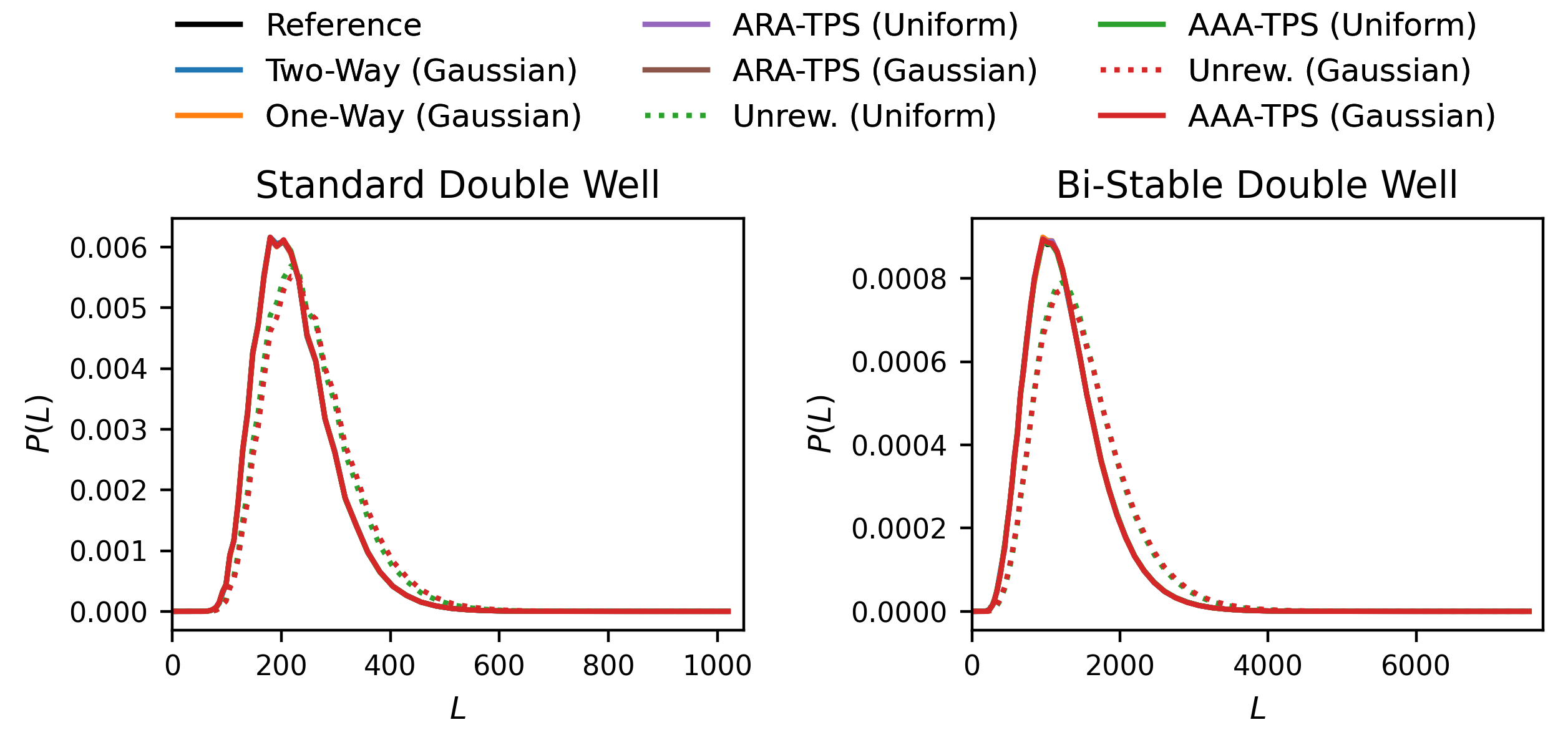}
    \caption{Distribution of path lengths using different TPS algorithms for the two-dimensional models presented in the main text. In addition to reference results (two-way TPS with uniform shooting point selection probability), state of the art methods (two-way and one-way with Gaussian selection probability) and the algorithms presented in this work (ARA-TPS and AAA-TPS with uniform and Gaussian selection probability) we also show the results obtained by always accepting paths without a posteriori reweighting (indicated as Unrew. in the legend). This highlight the presence of a bias generated by sampling the distribution $\widetilde{P}_{\mathrm{AB}}[X(L)]$. This bias favors longer paths with respect to the correct path distribution $P_{\mathrm{AB}}[X(L)]$. The reweighing procedure corrects for this bias.}
    \label{fig:main_pl}
\end{figure}

\subsection{Convergence of $p(x|\text{TP})$}
In the main text, we use a standard double well to prove the convergence of the proposed algorithm to the correct $p(x|\text{TP})$ of the given reference. The reference is obtained by performing 250 replicas of a two-way shooting TPS run with uniform shooting point selection probability made by $10^5$ sampling trials run after $2.5\times10^3$ equilibration trials. The curves in Fig.~2C of the main text are obtained performing 50 replicas with the same amount of trials and equilibration steps used for the reference for each of the algorithm presented.

\begin{figure}[htbp]
    \centering
    \includegraphics[width=\linewidth]{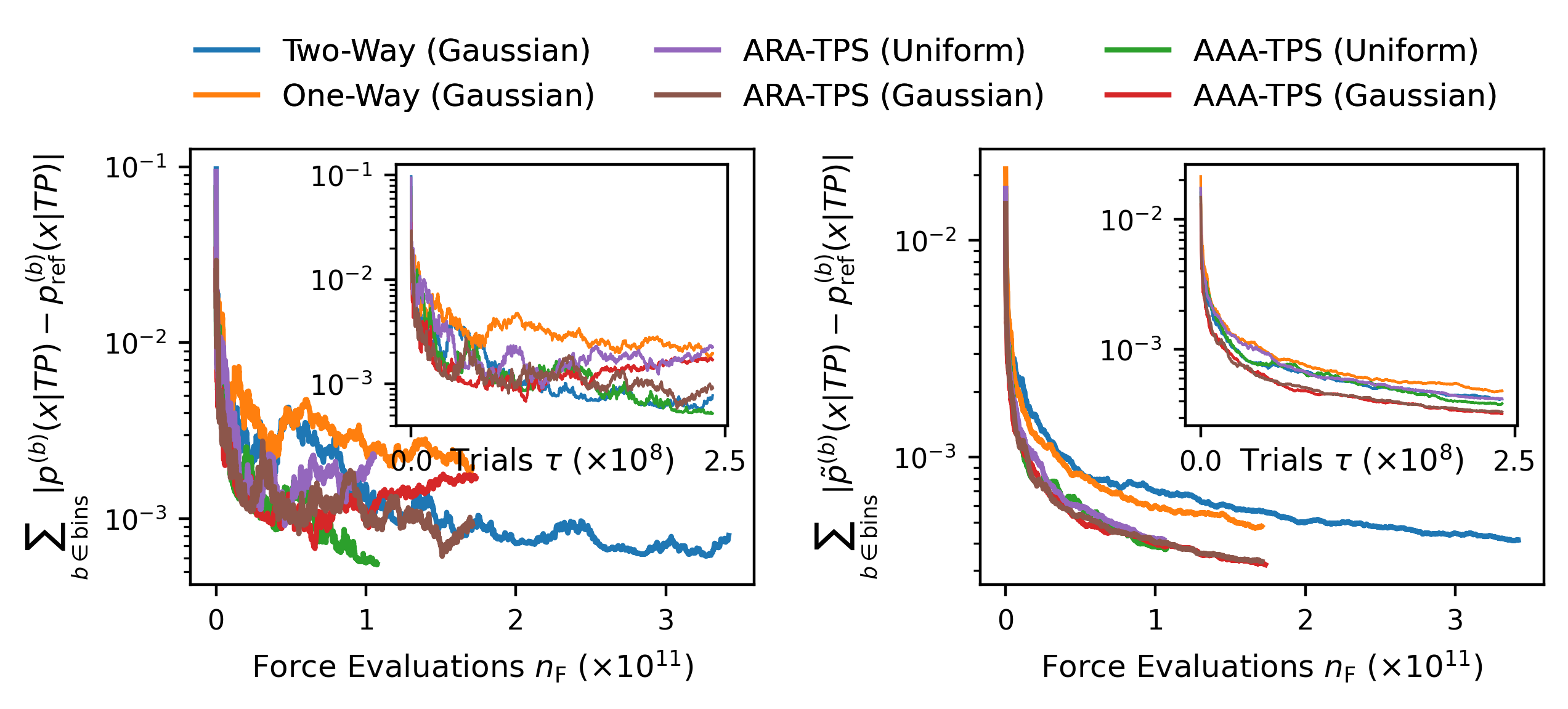}
    \caption{Convergence to reference path ensemble artificially symmetrized on $x^{(1)}$ of different TPS methods for a bi-stable double well. Symmetrization is performed by by adding to the ensemble the reflected along $x^{(1)}$ of each path sampled. Convergence is slower than in the standard double well (Fig.~2A of the main text) due to the correlations induced by the Markov chain, which tends to sample subsequent paths in the same transition channel (left panel). Once the number of paths in the two channel is artificially equalized (right panel), the convergence to the reference ensemble shows the same trend as in the case of a standard double well.}
    \label{fig:main_convergence}
\end{figure}

The same kind of analysis is much more challenging in the case of the bi-stable double well due to the tendency of TPS algorithms to get trapped in one of the transition channels of the system that connect $\mathrm{A}$ and $\mathrm{B}$. We show in Fig.~\ref{fig:main_convergence} the results of the convergence test for the bi-stable double well. The reference in this case is obtained performing 24 675 replicas of a two-way shooting TPS with uniform selection probability, again made by $10^5$ sampling trials run after $2.5\times10^3$ equilibration trials. In this case, to increase the accuracy of the of the reference, we manually symmetrize the histogram to have exactly a 1:1 ratio of paths in the upper and lower channel. The curves in Fig.~\ref{fig:main_convergence} are obtained performing 2 400 replicas with the same amount of trials and equilibration steps used for the reference for each of the algorithm presented. In the left panel we show the results obtained from the convergence analysis. The right panel shows the same results if we also artificially symmetrize the path distribution of the results, to highlight that the apparent non-convergence of the left plot is only to be attributed to correlation of samples and not to fallacies in the algorithm.

\subsection{Overdamped Regime of CO$_2$ Hydrate Formation}
The algorithm introduced in Sec.~II of the main text requires the process under investigation to be described by overdamped dynamics. 
It is therefore necessary to justify that for the system at hand there exists a relevant timescale at which overdamped dynamics will fully recover the dynamics of the full (non-overdamped) Langevin system and that this timescale is the one at which the formation of clathrates occurs. 

To do this, we assume that there exists a timescale $\tau_v$ at which the velocities have fully decorrelated. 
Given an initial time $\tau_0$, by $\tau_0 + \tau_v$ all memory of the initial velocity state is erased, and any observable will behave as if velocities are drawn freshly from the Boltzmann distribution.

In dense, relatively hot systems, frequent particle collisions cause rapid velocity decorrelation, leading to a short $\tau_v$. On the other hand, with particles being trapped in their dense environment, positional diffusion is severely restricted, leading to much slower positional decorrelation with long $\tau_r$.
For observables whose variations are significant only on timescales larger than $\tau_v$, it is therefore justified to collapse the full phase space $\Gamma = \{x_i, v_i\}$ into a restricted, position only phase space $\Gamma^* = \{x_i\}$.

To test this hypothesis, we calculate the normalized autocorrelation function $\langle x_j(0)x_j(\tau)\rangle$ of the velocity $v$ and the position $r$ as:
\begin{equation}
    \langle x_j(0)x_j(\tau)\rangle \approx \frac{1}{T_\mathrm{max}} \int_0^
    {T_\mathrm{max}} \frac{(x_j(t)-\bar{x}_j)(x_j(t+\tau)-\bar{x}_j)}
    {\sigma^2_{xj}} dt,
\end{equation}
where $T_\mathrm{max}$ is the observation time, $\bar{x}_j$ the equilibrium average of the variable $x$ for a particle $j$ and $\sigma^2_{xj}$ is the variance. This quantity is then averaged over all particles $j$ to obtain the particle-averaged result.

For a clathrate hydrate system the velocity and position autocorrelation functions are given also  for different stages along the crystallization transition. On the timescale of picoseconds (small enough to resolve in fine detail the clathrate formation process), the velocity autocorrelation function almost immediately decays to zero, indicating that the velocities are fully decorrelated. The position autocorrelation function, however, shows a much slower decay. This means that, at these timescales, the process of interested happens in the overdamped regime. When producing trajectories, is it then a good compromise between resolution and memory efficiency to save configurations with an interval (e.g. \SI{200}{\pico \second}) at which velocities have long decorrelated, while the positions are still highly correlated.

\begin{figure}[p]
    \centering
    \includegraphics[width=\linewidth]{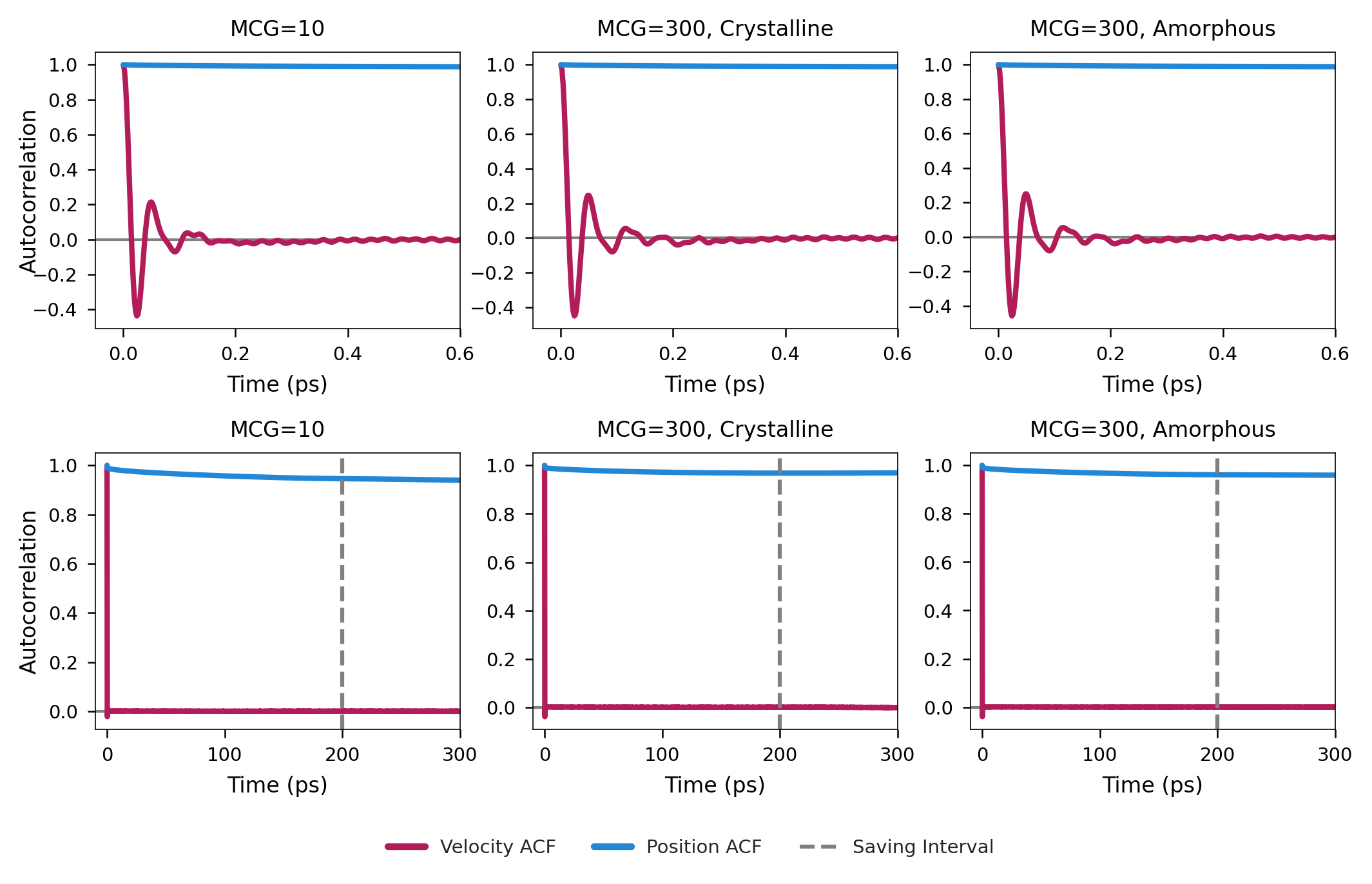}
    \caption{Autocorrelation functions (ACF) of the velocity and position of a clathrate hydrate system at different stages along the crystallization transition. The top panel shows the relaxation on a short timescale of up to \SI{0.6}{\pico \second} and the bottom panel shows the decorrelation on a longer timescale of up to \SI{300}{\pico \second}. The gray, dashed vertical line suggests a saving interval of 200 \si{ps} at which the velocities are fully decorrelated while the positions are still highly correlated.}
    \label{fig:ACFS}
\end{figure}

\subsection{Comparison of cage densities}
\begin{figure}[p]
    \centering
    \includegraphics[width=\linewidth]{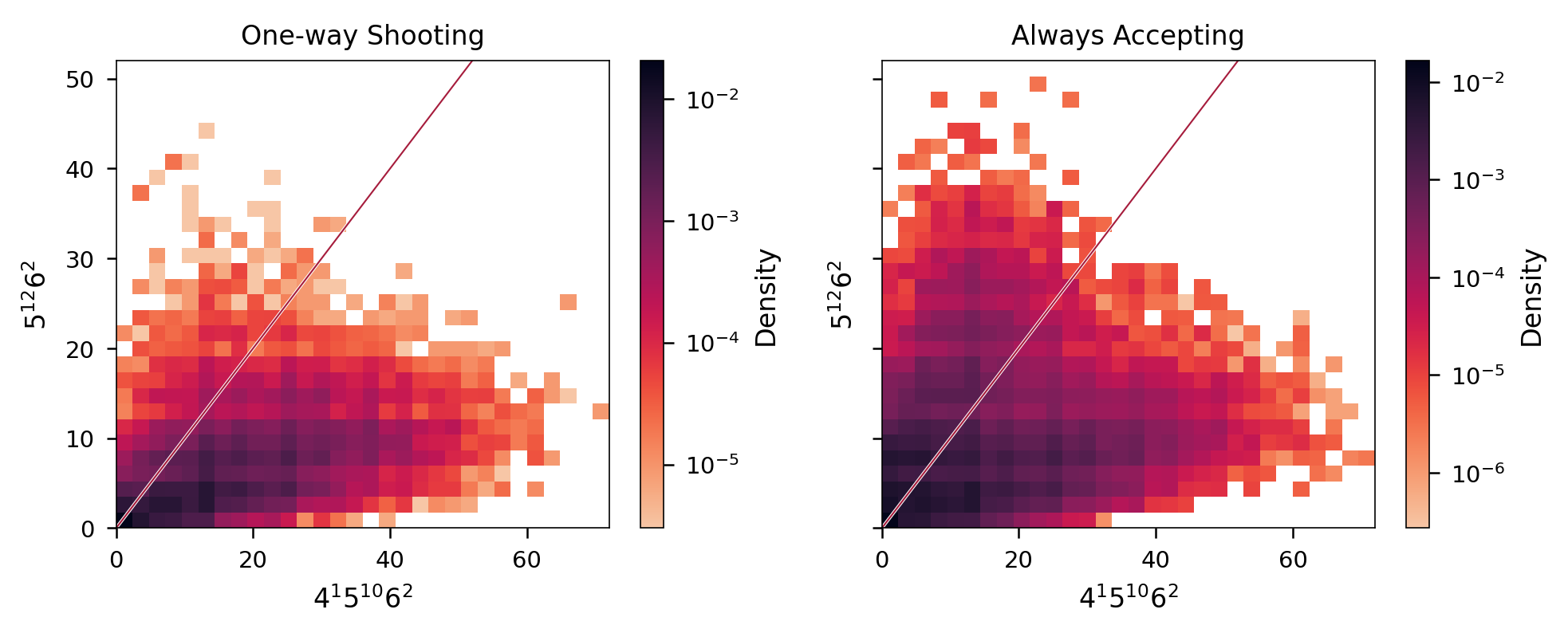}
    \caption{Histogram of the cage count densities of the $5^{12}6^2$ and $4^15^{10}6^2$ cages for the whole transition paths. The left panel shows the cage count density for one-way shooting, while the right panel shows the cage count density for AAA-TPS. The red line indicates an equal number of both cage types.}
    \label{fig:structure_analysis_histogram}
\end{figure}

In Fig.~\ref{fig:structure_analysis_histogram}, we show a comparison between cage populations along the sampled transition paths using standard one-way shooting and AAA-TPS. In the former case, the crystalline channel is much less explored than in the former, providing a transition path ensemble that is biased due to low sampling. Note that the left panel of Fig.~\ref{fig:structure_analysis_histogram} is very close to the results obtained in the upper left panel of Fig.~2 in Ref.~\citenum{sm_arjun_molecular_2021} where one-way shooting TPS has been performed on the same system with the same simulation set up. 

\end{document}